\newif\iflanl
\openin 1 lanlmac
\ifeof 1 \lanlfalse \else \lanltrue \fi
\closein 1
\iflanl
    \input lanlmac
\else
    \message{[lanlmac not found - use harvmac instead}
    \input harvmac
\fi
\newif\ifhypertex
\ifx\hyperdef\UnDeFiNeD
    \hypertexfalse
    \message{[HYPERTEX MODE OFF}
    \def\hyperref#1#2#3#4{#4}
    \def\hyperdef#1#2#3#4{#4}
    \def\hypernoname{}
    \def\e@tf@ur#1{}
    \def\eprt#1{{\tt #1}}
    \def\CERN{\address{CERN, CH--1211 Geneva 23, Switzerland}}
    \def\wl{W.\ Lerche}
    \def\sts{S.\ Stieberger}
\else
    \hypertextrue
    \message{[HYPERTEX MODE ON}
\def\eprt#1{{\tt
#1}}
\def\CERN{\address{

Theory Division, CERN, Geneva, Switzerland}}
\def\wl{

 W.\ Lerche}
\def\sts{
S.\ Stieberger}
\fi
\newif\ifdraft

\noblackbox
\catcode`\@=11
\newif\iffrontpage
\ifx\answ\bigans
\def\titleft{\titla}
\magnification=1200\baselineskip=14pt plus 2pt minus 1pt
%
\advance\hoffset by-0.075truein
\advance\voffset by1.truecm
\hsize=6.15truein\vsize=600.truept\hsbody=\hsize\hstitle=\hsize
\else\let\lr=L
\def\titleft{\titla}
\magnification=1000\baselineskip=14pt plus 2pt minus 1pt
%
\hoffset=-0.75truein\voffset=-.0truein
\vsize=6.5truein
\hstitle=8.truein\hsbody=4.75truein
\fullhsize=10truein\hsize=\hsbody
\fi
\parskip=4pt plus 15pt minus 1pt
%
\newif\iffigureexists
\newif\ifepsfloaded
\def\epsfcheck{
\ifdraft
\input epsf\epsfloadedtrue
\else
  \openin 1 epsf
  \ifeof 1 \epsfloadedfalse \else \epsfloadedtrue \fi
  \closein 1
  \ifepsfloaded
    \input epsf
  \else
\immediate\write20{NO EPSF FILE --- FIGURES WILL BE IGNORED}
  \fi
\fi
\def\epsfcheck{}}
\def\checkex#1{
\ifdraft
\figureexistsfalse\immediate%
\write20{Draftmode: figure #1 not included}
\figureexiststrue
\else\relax
    \ifepsfloaded \openin 1 #1
        \ifeof 1
           \figureexistsfalse
  \immediate\write20{FIGURE FILE #1 NOT FOUND}
        \else \figureexiststrue
        \fi \closein 1
    \else \figureexistsfalse
    \fi
\fi}
\def\missbox#1#2{$\vcenter{\hrule
\hbox{\vrule height#1\kern1.truein
\raise.5truein\hbox{#2} \kern1.truein \vrule} \hrule}$}
\def\lfig#1{
\let\labelflag=#1%
\def\numb@rone{#1}%
\ifx\labelflag\UnDeFiNeD%
{\xdef#1{\the\figno}%
\writedef{#1\leftbracket{\the\figno}}%
\global\advance\figno by1%
}\fi{\hyperref{}{figure}{{\numb@rone}}{Fig.{\numb@rone}}}}
\def\figinsert#1#2#3#4{
\epsfcheck\checkex{#4}%
\def\figsize{#3}%
\let\flag=#1\ifx\flag\UnDeFiNeD
{\xdef#1{\the\figno}%
\writedef{#1\leftbracket{\the\figno}}%
\global\advance\figno by1%
}\fi
\goodbreak\midinsert%
\iffigureexists
\centerline{\epsfysize\figsize\epsfbox{#4}}%
\else%
\vskip.05truein
  \ifepsfloaded
  \ifdraft
  \centerline{\missbox\figsize{Draftmode: #4 not included}}%
  \else
  \centerline{\missbox\figsize{#4 not found}}
  \fi
  \else
  \centerline{\missbox\figsize{epsf.tex not found}}
  \fi
\vskip.05truein
\fi%
{\smallskip%
\leftskip 4pc \rightskip 4pc%
\noindent\ninepoint\sl \baselineskip=11pt%
{\bf{\hyperdef\hypernoname{figure}{{#1}}{Fig.{#1}}}:~}#2%
\smallskip}\bigskip\endinsert%
}

\def\boxit#1{\vbox{\hrule\hbox{\vrule\kern8pt
\vbox{\hbox{\kern8pt}\hbox{\vbox{#1}}\hbox{\kern8pt}}
\kern8pt\vrule}\hrule}}
\def\mathboxit#1{\vbox{\hrule\hbox{\vrule\kern8pt\vbox{\kern8pt
\hbox{$\displaystyle #1$}\kern8pt}\kern8pt\vrule}\hrule}}
%
\font\bigit=cmti10 scaled \magstep1

\font\titla=cmr10 scaled\magstep3
\font\tenmss=cmss10
\font\absmss=cmss10 scaled\magstep1

\newfam\mssfam
\font\footrm=cmr8  \font\footrms=cmr5
\font\footrmss=cmr5   \font\footi=cmmi8
\font\footis=cmmi5   \font\footiss=cmmi5
\font\footsy=cmsy8   \font\footsys=cmsy5
\font\footsyss=cmsy5   \font\footbf=cmbx8
\font\footmss=cmss8
\def\footfont{\def\rm{\fam0\footrm}
\textfont0=\footrm \scriptfont0=\footrms
\scriptscriptfont0=\footrmss
\textfont1=\footi \scriptfont1=\footis
\scriptscriptfont1=\footiss
\textfont2=\footsy \scriptfont2=\footsys
\scriptscriptfont2=\footsyss
\textfont\itfam=\footi \def\it{\fam\itfam\footi}
\textfont\mssfam=\footmss \def\mss{\fam\mssfam\footmss}
\textfont\bffam=\footbf \def\bf{\fam\bffam\footbf} \rm}
\def\tenpoint{\def\rm{\fam0\tenrm}
\textfont0=\tenrm \scriptfont0=\sevenrm
\scriptscriptfont0=\fiverm
\textfont1=\teni  \scriptfont1=\seveni
\scriptscriptfont1=\fivei
\textfont2=\tensy \scriptfont2=\sevensy
\scriptscriptfont2=\fivesy
\textfont\itfam=\tenit \def\it{\fam\itfam\tenit}
\textfont\mssfam=\tenmss \def\mss{\fam\mssfam\tenmss}
\textfont\bffam=\tenbf \def\bf{\fam\bffam\tenbf} \rm}
\ifx\answ\bigans\def\abstractfont{\tenpoint}\else
\def\abstractfont{\def\rm{\fam0\absrm}
\textfont0=\absrm \scriptfont0=\absrms
\scriptscriptfont0=\absrmss
\textfont1=\absi \scriptfont1=\absis
\scriptscriptfont1=\absiss
\textfont2=\abssy \scriptfont2=\abssys
\scriptscriptfont2=\abssyss
\textfont\itfam=\bigit \def\it{\fam\itfam\bigit}
\textfont\mssfam=\absmss \def\mss{\fam\mssfam\absmss}
\textfont\bffam=\absbf \def\bf{\fam\bffam\absbf}\rm}\fi
%
\def\f@@t{\baselineskip10pt\lineskip0pt\lineskiplimit0pt
\bgroup\aftergroup\@foot\let\next}
\setbox\strutbox=\hbox{\vrule height 8.pt depth 3.5pt width\z@}
\def\vfootnote#1{\insert\footins\bgroup
\baselineskip10pt\footfont
\interlinepenalty=\interfootnotelinepenalty
\floatingpenalty=20000
\splittopskip=\ht\strutbox \boxmaxdepth=\dp\strutbox
\leftskip=24pt \rightskip=\z@skip
\parindent=12pt \parfillskip=0pt plus 1fil
\spaceskip=\z@skip \xspaceskip=\z@skip
\Textindent{$#1$}\footstrut\futurelet\next\fo@t}
\def\Textindent#1{\noindent\llap{#1\enspace}\ignorespaces}
\def\foot{\global\advance\ftno by1%
\attach{\hyperref{}{footnote}{\the\ftno}{\footsymbolgen}}%
\vfootnote{\hyperdef\hypernoname{footnote}{\the\ftno}{\footsymbol}}}%
\def\footnote#1{\global\advance\ftno by1%
\attach{\hyperref{}{footnote}{\the\ftno}{#1}}%
\vfootnote{\hyperdef\hypernoname{footnote}{\the\ftno}{#1}}}%
\newcount\lastf@@t           \lastf@@t=-1
\newcount\footsymbolcount    \footsymbolcount=0
\global\newcount\ftno \global\ftno=0
\def\footsymbolgen{\relax\footsym
\global\lastf@@t=\pageno\footsymbol}
\def\footsym{\ifnum\footsymbolcount<0
\global\footsymbolcount=0\fi
{\iffrontpage \else \advance\lastf@@t by 1 \fi
\ifnum\lastf@@t<\pageno \global\footsymbolcount=0
\else \global\advance\footsymbolcount by 1 \fi }
\ifcase\footsymbolcount
\fd@f\dagger\or \fd@f\diamond\or \fd@f\ddagger\or
\fd@f\natural\or \fd@f\ast\or \fd@f\bullet\or
\fd@f\star\or \fd@f\nabla\else \fd@f\dagger
\global\footsymbolcount=0 \fi }
\def\fd@f#1{\xdef\footsymbol{#1}}
\def\space@ver#1{\let\@sf=\empty \ifmmode #1\else \ifhmode
\edef\@sf{\spacefactor=\the\spacefactor}
\unskip${}#1$\relax\fi\fi}
\def\attach#1{\space@ver{\strut^{\mkern 2mu #1}}\@sf}
%
\newif\ifnref
\def\rrr#1#2{\relax\ifnref\nref#1{#2}\else\ref#1{#2}\fi}
\def\ldf#1#2{\begingroup\obeylines
\gdef#1{\rrr{#1}{#2}}\endgroup\unskip}
\def\nrf#1{\nreftrue{#1}\nreffalse}
\def\doubref#1#2{\refs{{#1},{#2}}}
\def\multref#1#2#3{\nrf{#1#2#3}\refs{#1{--}#3}}
\nreffalse
\def\refout{\listrefs}

\def\lref{\ldf}

\def\eqn#1{\xdef #1{(\noexpand\hyperref{}%
{equation}{\secsym\the\meqno}%
{\secsym\the\meqno})}\eqno(\hyperdef\hypernoname{equation}%
{\secsym\the\meqno}{\secsym\the\meqno})\eqlabeL#1%
\writedef{#1\leftbracket#1}\global\advance\meqno by1}
\def\eqnalign#1{\xdef #1{\noexpand\hyperref{}{equation}%
{\secsym\the\meqno}{(\secsym\the\meqno)}}%
\writedef{#1\leftbracket#1}%
\hyperdef\hypernoname{equation}%
{\secsym\the\meqno}{\e@tf@ur#1}\eqlabeL{#1}%
\global\advance\meqno by1}
\def\eqnalign#1{\xdef #1{(\secsym\the\meqno)}
\writedef{#1\leftbracket#1}%
\global\advance\meqno by1 #1\eqlabeL{#1}}
%

%
\def\chap#1{\newsec{#1}}
\def\chapter#1{\chap{#1}}
\def\sect#1{\subsec{#1}}
\def\section#1{\sect{#1}}
\def\\{\ifnum\lastpenalty=-10000\relax
\else\hfil\penalty-10000\fi\ignorespaces}
\def\note#1{\leavevmode%
\edef\@@marginsf{\spacefactor=\the\spacefactor\relax}%
\ifdraft\strut\vadjust{%
\hbox to0pt{\hskip\hsize%
\ifx\answ\bigans\hskip.1in\else\hskip .1in\fi%
\vbox to0pt{\vskip-\dp
\strutbox\sevenbf\baselineskip=8pt plus 1pt minus 1pt%
\ifx\answ\bigans\hsize=.7in\else\hsize=.35in\fi%
\tolerance=5000 \hbadness=5000%
\leftskip=0pt \rightskip=0pt \everypar={}%
\raggedright\parskip=0pt \parindent=0pt%
\vskip-\ht\strutbox\noindent\strut#1\par%
\vss}\hss}}\fi\@@marginsf\kern-.01cm}
\def\titlepage{%
\frontpagetrue\nopagenumbers\abstractfont%
\hsize=\hstitle\rightline{\vbox{\baselineskip=10pt%
{\abstractfont\pubnum}}}\pageno=0}
\frontpagefalse
\def\pubnum{}
\def\pdate{\number\month/\number\yearltd}
\def\makefootline{\iffrontpage\vskip .27truein
\line{\the\footline}
\vskip -.1truein\leftline{\vbox{\baselineskip=10pt%
{\abstractfont\pdate}}}
\else\vskip.5cm\line{\hss \tenrm $-$ \folio\ $-$ \hss}\fi}
\def\title#1{\vskip .7truecm\titlestyle{\titleft #1}}
\def\titlestyle#1{\par\begingroup \interlinepenalty=9999
\leftskip=0.02\hsize plus 0.23\hsize minus 0.02\hsize
\rightskip=\leftskip \parfillskip=0pt
\hyphenpenalty=9000 \exhyphenpenalty=9000
\tolerance=9999 \pretolerance=9000
\spaceskip=0.333em \xspaceskip=0.5em
\noindent #1\par\endgroup }
\def\autskip{\ifx\answ\bigans\vskip.5truecm\else\vskip.1cm\fi}
\def\author#1{\vskip .7in \centerline{#1}}

\def\address#1{\ifx\answ\bigans\vskip.2truecm
\else\vskip.1cm\fi{\it \centerline{#1}}}
\def\abstract#1{
\vskip .5in\vfil\centerline
{\bf Abstract}\penalty1000
{{\smallskip\ifx\answ\bigans\leftskip 2pc \rightskip 2pc
\else\leftskip 5pc \rightskip 5pc\fi
\noindent\abstractfont \baselineskip=12pt
{#1} \smallskip}}
\penalty-1000}
\def\endpage{\tenpoint\supereject\global\hsize=\hsbody%
\frontpagefalse\footline={\hss\tenrm\folio\hss}}
\def\ack{\goodbreak\vskip2.cm\centerline{{\bf Acknowledgements}}}
%
%

%
\def\bfone{\relax{\rm 1\kern-.35em 1}}
\def\inbar{\vrule height1.5ex width.4pt depth0pt}
\def\IC{\relax\,\hbox{$\inbar\kern-.3em{\mss C}$}}
\def\ID{\relax{\rm I\kern-.18em D}}
\def\IF{\relax{\rm I\kern-.18em F}}
\def\IH{\relax{\rm I\kern-.18em H}}
\def\II{\relax{\rm I\kern-.17em I}}
\def\IN{\relax{\rm I\kern-.18em N}}
\def\IP{\relax{\rm I\kern-.18em P}}
\def\IQ{\relax\,\hbox{$\inbar\kern-.3em{\rm Q}$}}
\def\IR{\relax{\rm I\kern-.18em R}}
\font\cmss=cmss10 \font\cmsss=cmss10 at 7pt
\def\ZZ{\relax\ifmmode\mathchoice
{\hbox{\cmss Z\kern-.4em Z}}{\hbox{\cmss Z\kern-.4em Z}}
{\lower.9pt\hbox{\cmsss Z\kern-.4em Z}}
{\lower1.2pt\hbox{\cmsss Z\kern-.4em Z}}\else{\cmss Z\kern-.4em
Z}\fi}

\def\a{\alpha}  
 \def\c{\gamma}
 \def\l{\lambda}
 
\def\cA{{\cal A}} 
 
\def\cE{{\cal E}}
\def\cF{{\cal F}} \def\cG{{\cal G}}
\def\cH{{\cal H}} 
 
 \def\cM{{\cal M}}

\def\nup#1({Nucl.\ Phys.\ $\us {B#1}$\ (}
\def\plt#1({Phys.\ Lett.\ $\us  {#1}$\ (}
\def\cmp#1({Comm.\ Math.\ Phys.\ $\us  {#1}$\ (}
\def\prp#1({Phys.\ Rep.\ $\us  {#1}$\ (}
\def\prl#1({Phys.\ Rev.\ Lett.\ $\us  {#1}$\ (}
\def\prv#1({Phys.\ Rev.\ $\us  {#1}$\ (}
\def\mpl#1({Mod.\ Phys.\ Let.\ $\us  {A#1}$\ (}
\def\ijmp#1({Int.\ J.\ Mod.\ Phys.\ $\us{A#1}$\ (}
\def\tit#1|{{\it #1},\ }
%

%

\def\ni{\noindent}
\def\tilde{\widetilde}
\def\bar{\overline}
\def\us#1{\underline{#1}}

\def\hat{\widehat}

\def\Coeff#1#2{{#1\over #2}}
\def\Coe#1.#2.{{#1\over #2}}
\def\coeff#1#2{\relax{\textstyle {#1 \over #2}}\displaystyle}
\def\coe#1.#2.{\relax{\textstyle {#1 \over #2}}\displaystyle}

\def\shalf{\relax{\textstyle {1 \over 2}}\displaystyle}

\def\to{\rightarrow}
\def\notin{\hbox{{$\in$}\kern-.51em\hbox{/}}}

\def\Trbel#1{\mathop{{\rm Tr}}_{#1}}

\def\attac#1{\Bigl\vert
{\phantom{X}\atop{{\rm\scriptstyle #1}}\phantom{X}}}

\def\del{\partial}

\def\nex#1{$N\!=\!#1$}


\catcode`\@=12

\def\h {{1\over 2}}
\def\al {\alpha}

\def\ov {\overline}
\def\o {\over}
\def\Li {{\cal L}i}

\def\E {\hat E}
\def\th {\theta}

\def\tr {{\rm Tr}}

\def\lf {\left}
\def\ri {\right}

\def\al {\alpha}
\def\re {{\rm Re}}
\def\im {{\rm Im}}
\def\p {\partial}

\def\lf {\Big}
\def\ri {\Big}

\def\appA{A}
\def\appB{B}

\def\het{{\rm het}}
\def\Sh{S_\het}

\def\st{^{{\rm (st)}}}
\def\kt{{(K3)}}
\def\attac#1{\Bigl\vert
{\phantom{X}\atop{{\scriptstyle #1}}\phantom{X}}}

 1
\def\CHI{{\cE}}

\def\smsup#1{{\scriptscriptstyle{#1}}}

\def\cFT{{\cF_1^{\smsup{(T^2)}}}}

\def\fk{{\smsup{[2K\!+4]}}}

\def\nihil#1{{\sl #1}}
\def\br{\hfill\break}
\def\np {{ Nucl.\ Phys.} {\bf B}}

\def\m{{\bf m}}
\def\n{{\bf n}}

\def\IZ{\ZZ}
\def\DD{{\tilde {\cal A}}}


\def\nihil#1{{\sl #1}}
\def\br{\hfill\break}
\def\np {{ Nucl.\ Phys.} {\bf B}}

\def\ijmp {{Int. J. Mod. Phys.\ }{\bf A}}

\lref\elias{
{
E.\ Kiritsis,
\nihil{Introduction to nonperturbative string theory,}
\eprt{hep-th/9708130};
}
{
\nihil{Duality and instantons in string theory,}
\eprt{hep-th/9906018}.
}}

\lref\KK{E.\ Kiritsis and C.\ Kounnas,
\nihil{Perturbative and nonperturbative partial
supersymmetry breaking: N=4 $\to$ N=2 $\to$ N=1,}
 Nucl.\ Phys.\ {\bf B503} 117 (1997),
\eprt{hep-th/9703059}.
}

\lref\wilson{
{G.\ Lopes Cardoso, G.\ Curio and D.\ L\"ust,
\nihil{Perturbative couplings and
modular forms in N=2 string models with a Wilson line,}
 Nucl.\ Phys.\ {\bf B491} 147 (1997),
\eprt{hep-th/9608154};
}
\br
{S.\ Stieberger,
\nihil{(0,2) heterotic gauge couplings and their M theory origin,}
 Nucl.\ Phys.\ {\bf B541} 109 (1999),
\eprt{hep-th/9807124}.
}}

\lref\fklz{H. Ooguri and C. Vafa, {\it Geometry of N=2 strings},
\np {\bf 361} (1991) 469;\br
S. Ferrara, C. Kounnas, D. L\"ust and F. Zwirner, {\it Duality
invariant partition functions and automorphic superpotentials
for (2,2) string compactifications}, \np {\bf 365} (1991) 431.}

\lref\MKPCR{R.\ de Mello Koch, A.\ Paulin-Campbell and J.P.\ Rodrigues,
\nihil{Nonholomorphic corrections from three-branes in F theory,}
\eprt{hep-th/9903029}.
}

\lref\igusa{J.\ Igusa, Amer.\ J.\ Math.\ 84 (1962) 175; 86 (1964) 392.}

\lref\BK{C.\ Bachas and E.\ Kiritsis,
\nihil{$F^4$ terms in N=4 string vacua,}
 Nucl.\ Phys.\ Proc.\ Suppl.\ {\bf 55B} 194 (1997),
\eprt{hep-th/9611205}.
}

\lref\LS{
{W.\ Lerche and S.\ Stieberger,
\nihil{Prepotential, mirror map and F theory on K3,}
 Adv.\ Theor.\ Math.\ Phys.\ {\bf 2} 1105 (1998),
\eprt{hep-th/9804176};
}
\br
{W.\ Lerche, S.\ Stieberger and N.P.\ Warner,
\nihil{Quartic gauge couplings from K3 geometry,}
\eprt{hep-th/9811228};
}
{
\nihil{Prepotentials from symmetric products,}
\eprt{hep-th/9901162}.
}
}

\lref\DS{M.\ Dine and N.\ Seiberg,
\nihil{Comments on higher derivative operators in some SUSY field
theories,}
 Phys.\ Lett.\ {\bf B409} 239 (1997),
\eprt{hep-th/9705057}.
}

\lref\seiberg{N.\ Seiberg,
\nihil{Notes on theories with 16 supercharges,}
 Nucl.\ Phys.\ Proc.\ Suppl.\ {\bf 67} 158 (1998),
\eprt{hep-th/9705117}.
}

\lref\harveymoore{J.A.\ Harvey and G.\ Moore,
\nihil{Five-brane instantons and $R^2$ couplings in N=4 string theory,}
 Phys.\ Rev.\ {\bf D57} 2323 (1998),
\eprt{hep-th/9610237}.
}

\lref\greekRtwo{A.\ Gregori, E.\ Kiritsis,
C.\ Kounnas, N.A.\ Obers,
P.M.\ Petropoulos and B.\ Pioline,
\nihil{$R^2$ corrections and nonperturbative
dualities of N = 4 string ground states,}
 Nucl.\ Phys.\ {\bf B510} 423 (1998),
\eprt{hep-th/9708062}.
}

\lref\MMS{J.\ Maldacena, G.\ Moore and A.\ Strominger,
\nihil{Counting BPS black holes in toroidal type-II string theory,}
\eprt{hep-th/9903163}.
}

\lref\HM{J.A.\ Harvey and G.\ Moore,
\nihil{Algebras, BPS states, and strings,}
 Nucl.\ Phys.\ {\bf B463} 315 (1996),
\eprt{hep-th/9510182}.
}

\lref\WL{W.\ Lerche,
\nihil{Elliptic Index And Superstring Effective Actions,}
 Nucl.\ Phys.\ {\bf B308} 102 (1988).
}

\lref\ghosttrial{
{W.\ Lerche, A.N.\ Schellekens and N.P.\ Warner,
\nihil{Ghost Triality And Superstring Partition Functions,}
 Phys.\ Lett.\ {\bf B214} 41 (1988);
}
{
\nihil{Lattices And Strings,}
 Phys.\ Rept.\ {\bf 177} 1 (1989).
}
}

\lref\KthreeBPS{E.\ Witten,
\nihil{Small instantons in string theory,}
 Nucl.\ Phys.\ {\bf B460} 541 (1996),
\eprt{hep-th/9511030}.
}

\lref\DVV{
{R.\ Dijkgraaf, E.\ Verlinde and H.\ Verlinde,
\nihil{Counting dyons in N=4 string theory,}
 Nucl.\ Phys.\ {\bf B484} 543 (1997),
\eprt{hep-th/9607026};
}\br
{R.\ Dijkgraaf, G.\ Moore, E.\ Verlinde and H.\ Verlinde,
\nihil{Elliptic genera of symmetric products and second quantized
strings,}
 Commun.\ Math.\ Phys.\ {\bf 185} 197 (1997),
\eprt{hep-th/9608096}.
}
}

\lref\dieterBPS{G.\ Lopes Cardoso, G.\ Curio, D.\ L\"ust,
T.\ Mohaupt and S.\ Rey,
\nihil{BPS spectra and nonperturbative gravitational
couplings in N=2, N=4 supersymmetric string theories,}
 Nucl.\ Phys.\ {\bf B464} 18 (1996),
\eprt{hep-th/9512129}.
}

\lref\AKEZ{A.\ Klemm and E.\ Zaslow,
\nihil{Local mirror symmetry at higher genus,}
\eprt{hep-th/9906046}.
}

\lref\GPZ{L.\ Girardello, M.\ Porrati and A.\ Zaffaroni,
\nihil{Heterotic Type II string duality and the H monopole problem,}
 Int.\ J.\ Mod.\ Phys.\ {\bf A11} 4255 (1996),
\eprt{hep-th/9508056}.
}

\lref\bcov{
{M.\ Bershadsky, S.\ Cecotti, H.\ Ooguri and C.\ Vafa,
\nihil{Holomorphic anomalies in topological field theories,}
Nucl.\ Phys.\ {\bf B405} 279 (1993),
\eprt{hep-th/9302103};
}
{
\nihil{Kodaira--Spencer theory of gravity and exact
results for quantum string amplitudes,}
Commun.\ Math.\ Phys.\ {\bf 165} 311 (1994),
\eprt{hep-th/9309140}.
}
}

\lref\StV{A.\ Strominger and C.\ Vafa,
\nihil{Microscopic origin of the Bekenstein-Hawking entropy,}
 Phys.\ Lett.\ {\bf B379} 99 (1996),
\eprt{hep-th/9601029}.
}

\lref\MDXL{M.J.\ Duff and J.X.\ Lu,
\nihil{Elementary five-brane solutions of D = 10 supergravity,}
 Nucl.\ Phys.\ {\bf B354} 141 (1991).
}

\lref\CHS{
{C.G.\ Callan, J.A.\ Harvey and A.\ Strominger,
\nihil{World sheet approach to heterotic instantons and solitons,}
 Nucl.\ Phys.\ {\bf B359} 611 (1991);
}
{
\nihil{Worldbrane actions for string solitons,}
 Nucl.\ Phys.\ {\bf B367} 60 (1991).
}
}

\lref\Gm{G.\ Moore,
\nihil{String duality, automorphic forms,
and generalized Kac-Moody algebras,}
 Nucl.\ Phys.\ Proc.\ Suppl.\ {\bf 67} 56 (1998),
\eprt{hep-th/9710198}.
}

\lref\SS{J.H.\ Schwarz and A.\ Sen,
\nihil{Duality symmetries of 4-D heterotic strings,}
 Phys.\ Lett.\ {\bf B312} 105 (1993),
\eprt{hep-th/9305185}.
}

\lref\onequarterBPS{M.J.\ Duff, J.T.\ Liu and J.\ Rahmfeld,
\nihil{Four-dimensional string-string-string triality,}
 Nucl.\ Phys.\ {\bf B459} 125 (1996),
\eprt{hep-th/9508094}.
}

\lref\GKPR{F.\ Gonzalez-Rey, B.\ Kulik, I.Y.\ Park and M.\ Rocek,
\nihil{Selfdual effective action of N=4 superYang-Mills,}
 Nucl.\ Phys.\ {\bf B544} 218 (1999),
\eprt{hep-th/9810152}.
}

\lref\GV{R.\ Gopakumar and C.\ Vafa,
\nihil{M-theory and topological strings.\ 1,2}
\eprt{hep-th/9809187}, \eprt{hep-th/9812127}.
}

\lref\BV{N.\ Berkovits and C.\ Vafa,
\nihil{N=4 topological strings,}
 Nucl.\ Phys.\ {\bf B433} 123 (1995),
\eprt{hep-th/9407190}.
}

\lref\FS{
{K.\ F\"orger and S.\ Stieberger,
\nihil{String amplitudes and N=2, d = 4 prepotential in
heterotic $K3\times T^2$ compactifications,}
 Nucl.\ Phys.\ {\bf B514} 135 (1998),
\eprt{hep-th/9709004};
}\br
{K.\ Foerger and S.\ Stieberger,
\nihil{Higher derivative couplings and heterotic
type I duality in eight dimensions,}
Nucl.\ Phys.\ {\bf B559} (1999) 277, \eprt{hep-th/9901020}.
}
}

\lref\JHGM{J.A.\ Harvey and G.\ Moore,
\nihil{Exact gravitational threshold correction in the FHSV model,}
 Phys.\ Rev.\ {\bf D57} 2329 (1998),
\eprt{hep-th/9611176}.
}

\lref\md{
{M.J.\ Duff and R.R.\ Khuri,
\nihil{Four-dimensional string/string duality,}
 Nucl.\ Phys.\ {\bf B411} 473 (1994),
\eprt{hep-th/9305142}.
}
{M.J.\ Duff,
\nihil{Strong/weak coupling duality from the dual string,}
 Nucl.\ Phys.\ {\bf B442} 47 (1995),
\eprt{hep-th/9501030}.
}
}

\lref\BVNfour{N.\ Berkovits and C.\ Vafa,
\nihil{N=4 topological strings,}
 Nucl.\ Phys.\ {\bf B433} 123 (1995),
\eprt{hep-th/9407190}.
}
\lref\BVRH{N.\ Berkovits and C.\ Vafa,
\nihil{Type IIB $R^4 H^{4g-4}$ conjectures,}
 Nucl.\ Phys.\ {\bf B533} 181 (1998),
\eprt{hep-th/9803145}.
}

\lref\SenII{A.\ Sen,
\nihil{String string duality conjecture in six-dimensions
and charged solitonic strings,}
 Nucl.\ Phys.\ {\bf B450} 103 (1995),
\eprt{hep-th/9504027}.
}

\lref\HS{J.A.\ Harvey and A.\ Strominger,
\nihil{The heterotic string is a soliton,}
 Nucl.\ Phys.\ {\bf B449} 535 (1995),
\eprt{hep-th/9504047}.
}

\lref\sen{A.\ Sen,
\nihil{Strong-weak coupling duality in
 four-dimensional string theory,}
 Int.\ J.\ Mod.\ Phys.\ {\bf A9} 3707 (1994),
\eprt{hep-th/9402002}.
}

\lref\FKP{S.\ Ferrara, C.\ Kounnas and M.\ Porrati,
\nihil{General Dimensional Reduction Of
Ten-Dimensional Supergravity And Superstring,}
 Phys.\ Lett.\ {\bf 181B} 263 (1986).
}

\lref \DKLII{L.J.\ Dixon, V.\ Kaplunovsky and J.\ Louis,
\nihil{Moduli dependence of string loop
corrections to gauge coupling constants,}
 Nucl.\ Phys.\ {\bf B355} 649 (1991).
}

\lref\msi{P.\ Mayr and S.\ Stieberger,
\nihil{Threshold corrections to gauge
couplings in orbifold compactifications,}
 Nucl.\ Phys.\ {\bf B407} 725 (1993),
\eprt{hep-th/9303017}.
}

\lref\fs{K.\ F\"orger and S.\ Stieberger,
\nihil{String amplitudes and N=2, d = 4
prepotential in heterotic $K3\times T^2$ compactifications,}
 Nucl.\ Phys.\ {\bf B514} 135 (1998),
\eprt{hep-th/9709004}.
}

\lref\BFKOV{
{C.\ Bachas, C.\ Fabre, E.\ Kiritsis,
 N.A.\ Obers and P.\ Vanhove,
\nihil{Heterotic / type I duality and D-brane instantons,}
 Nucl.\ Phys.\ {\bf B509} 33 (1998),
\eprt{hep-th/9707126};
}
{
E.\ Kiritsis and N.A.\ Obers,
\nihil{Heterotic type I duality
in $d <10$-dimensions, threshold corrections and D-instantons,}
 JHEP{\bf 10} 004 (1997),
\eprt{hep-th/9709058}.
}
}

\lref\kawai{
{T.\ Kawai,
\nihil{N=2 heterotic string threshold correction,
K3 surface and generalized Kac-Moody superalgebra,}
 Phys.\ Lett.\ {\bf B372} 59 (1996),
\eprt{hep-th/9512046};
}{
\nihil{String duality and enumeration of curves by Jacobi forms,}
\eprt{hep-th/9804014}.
}
}

\lref\stieberg{S. Stieberger,
\nihil{(0,2) heterotic gauge couplings and their M-theory origin},
Nucl.\ Phys.\ {\bf B541}, 109 (1999),
\eprt{hep-th/9807124}.
}

\lref\ms{P.\ Mayr and S.\ Stieberger,
\nihil{Moduli dependence of one loop gauge couplings in (0,2)
compactifications,}
 Phys.\ Lett.\ {\bf B355} 107 (1995),
\eprt{hep-th/9504129}.
}

\lref\network{A.\ Sen,
\nihil{String network,}
 JHEP {\bf 03} 005 (1998),
\eprt{hep-th/9711130}.
}

\lref\johnS{J.H.\ Schwarz,
\nihil{Lectures on superstring and M theory dualities: Given at ICTP
Spring School and at TASI Summer School,}
 Nucl.\ Phys.\ Proc.\ Suppl.\ {\bf 55B} 1 (1997),
\eprt{hep-th/9607201}.
}

\lref\ellg {A.\ Schellekens and N.\ Warner,
{\nihil{Anomalies, characters and strings,}
 Nucl.\  Phys.\ {\bf B287} (1987) 317;}\br
{E.\ Witten,
 \nihil{Elliptic genera and quantum field theory,}
 Commun.\  Math.\  Phys.\ {\bf 109} (1987) 525;}\br
W. Lerche, B.E.W. Nilsson, A.N. Schellekens and N.P. Warner,
\np {\bf 299} (1988) 91.}

\lref\mirror{
See e.g.,
\nihil{Essays on mirror manifolds}, (S.\ Yau, ed.),
International Press 1992;
\nihil{Mirror symmetry II}, (B.\ Greene et al, eds.),
International Press 1997.
}

\lref\MDGD{M.R.\ Douglas,
\nihil{Gauge fields and D-branes,}
 J.\ Geom.\ Phys.\ {\bf 28} 255 (1998),
\eprt{hep-th/9604198}.
}

\lref\MDBB{M.R.\ Douglas,
\nihil{Branes within branes,}
\eprt{hep-th/9512077}.
}

\lref\JM{J.\ Mourad,
\nihil{Anomalies of the SO(32) five-brane and their cancellation,}
 Nucl.\ Phys.\ {\bf B512} 199 (1998),
\eprt{hep-th/9709012}.
}

\lref\coniB{D.\ Ghoshal and C.\ Vafa,
\nihil{C = 1 string as the topological theory of the conifold,}
 Nucl.\ Phys.\ {\bf B453} 121 (1995),
\eprt{hep-th/9506122}.
}

\lref\coniA{C.\ Vafa,
\nihil{A Stringy test of the fate of the conifold,}
 Nucl.\ Phys.\ {\bf B447} 252 (1995),
\eprt{hep-th/9505023}.
}

\lref\AGNT{I.\ Antoniadis, E.\ Gava, K.S.\ Narain and T.R.\ Taylor,
\nihil{Topological amplitudes in string theory,}
Nucl.\ Phys.\ {\bf B413} 162 (1994),
\eprt{hep-th/9307158};\br
{M.\ Marino and G.\ Moore,
 \nihil{Counting higher genus curves in a Calabi-Yau manifold,}
Nucl.Phys.\ {\bf B543} 592 (1999),  \eprt{hep-th/9808131}.}}

\lref\soojong{S.\ Rey,
\nihil{The confining phase of superstrings and axionic strings,}
 Phys.\ Rev.\ {\bf D43} 526 (1991).
}

\lref\senKK{A.\ Sen,
\nihil{Kaluza-Klein dyons in string theory,}
 Phys.\ Rev.\ Lett.\ {\bf 79} 1619 (1997),
\eprt{hep-th/9705212}.
}

\lref\minahan{
{J.A.\ Minahan,
\nihil{One Loop Amplitudes On Orbifolds And
The Renormalization Of Coupling Constants,}
 Nucl.\ Phys.\ {\bf B298} 36 (1988);
}
\br
{P.\ Mayr and S.\ Stieberger,
\nihil{Dilaton, antisymmetric tensor and
gauge fields in string effective theories at the one loop level,}
 Nucl.\ Phys.\ {\bf B412} 502 (1994),
\eprt{hep-th/9304055};
}
\br
{K.\ F\"orger, B.A.\ Ovrut, S.J.\ Theisen and D.\ Waldram,
\nihil{Higher derivative gravity in string theory,}
 Phys.\ Lett.\ {\bf B388} 512 (1996),
\eprt{hep-th/9605145}.
}
}

\lref\contr{{
D.J. Gross, J.A. Harvey, E. Martinec and R. Rohm, \nihil{Heterotic
string theory (2). The interacting heterotic string}, Nucl.\ Phys. {\bf
B267} (1986) 75;\br W. Lerche, B.E.W Nilsson, A.N. Schellekens and N.P.
Warner, \nihil{Anomaly cancelling terms from the elliptic genus},
Nucl.\ Phys.  {\bf B299} (1988) 91;}\br {J. Ellis, P. Jetzer and L.
Mizrachi, \nihil{One--loop corrections to the effective field theory},
Nucl.\ Phys. {\bf B303} (1988) 1.}}

\lref\morales{
A.\ Hammou and J.\ Morales,
\nihil{Fivebrane instantons and higher derivative
couplings in type I theory,}
\eprt{hep-th/9910144}.
}



\def\pubnum{
\hbox{CERN-TH/99-210}
\hbox{hep-th/9907133}
\hbox{revised Oct 1999}}
\def\pdate{}
\titlepage
\vskip2.cm
\title
{{\titlefont 1/4 BPS States
and Non--Perturbative Couplings in N=4 String Theories}}
\vskip -.7cm
\autskip
\author{\wl\ and \sts}
\vskip0.2truecm
\CERN
\vskip-.4truecm

\abstract{
We compute certain $(2K\!+4)$--point, one--loop couplings in the type
IIA string compactified on $K3\times T^2$, which are related to a
topological index on this manifold. Their special feature is that they
are sensitive to only short and intermediate BPS multiplets.  The
couplings derive from underlying prepotentials $\cG^\fk(T,U)$, which
can be summed up to a generating function in the form:
$\sum_{K=0}^\infty  V^{2K}\!\!/(2K)!\,\,\cG^\fk(T,U) =\sum
d(4kl\!-\!m^2)\ \Li_{3}[e^{2\pi i(kT+lU)}e^{mV}]$. In the dual
heterotic string on $T^6$, the amplitudes describe non--perturbative
gravitational corrections to $K$-loop amplitudes due  to bound states
of fivebrane instantons with heterotic world-sheet instantons. We
argue, as a consequence, that our results also give information about
instanton configurations in six dimensional $Sp(2k)$ gauge theories
on~$T^6$.
}

\vfil
\vskip 1.cm
\ni {CERN-TH/99-210}\hfill\break
\ni July 1999
\endpage
\baselineskip=14pt plus 2pt minus 1pt


\chapter{Introduction}

BPS-saturated string loop amplitudes
\multref\coniA{\HM\BK\BFKOV\LS\GV}\elias\ play an important r\^ole
since they can give exact non-perturbative answers for appropriate dual
formulations of a given theory. The corresponding pieces of the
effective action are often given by holomorphic prepotentials, and it
is this holomorphicity which underlies their computability.
Particularly well-known are the couplings in \nex2 supersymmetric
string compactifications that describe gauge and certain gravitational
interactions. They are characterized by holomorphic prepotentials
$\cF_g$ \multref\bcov{\coniB\AGNT\JHGM}\AKEZ,  which can be
geometrically computed via mirror symmetry \mirror\ on Calabi-Yau
threefolds. An analogous holomorphic structure arises also in certain
eight dimensional string vacua \LS.

However, a comparable systematic treatment for four dimensional string
theories with more, notably \nex4 supersymmetries has been lacking so
far. The main novel feature in \nex4 supersymmetry is the appearance of
intermediate (``1/4 BPS'')  besides the short (``1/2 BPS'')
supermultiplets. An example for a 1/2 BPS saturated amplitude is given
by $\del_T\langle R^2\rangle$, which is perturbatively  exact at one
loop order in the type IIA string compactified on $K3\times T^2$. It
has been shown in \refs{\bcov{,}\harveymoore{,}\greekRtwo}\ to be given
by the $T$--derivative of:\foot{The regularization constant is
$\kappa\equiv {8\pi \o 3\sqrt 3}e^{1-\gamma_E}$, where $\gamma_E$ is
the Euler constant. In the type IIB string on $K3\times T^2$, it is the
$U$--derivative of this function what becomes relevant.}
$$
\eqalign{
\cF_1^{\scriptscriptstyle{(K3\!\times\!T^2)}}(T,U)\ &=\
\int{d^2\tau\over\tau_2}\Trbel{K3\times T^2}
\Big[(-1)^{J_R+J_L}J_RJ_Lq^{L_0}\bar q^{\bar L_0}-24\Big]
\cr
&=\ 24 \Big[\ln(T_2|\eta(T)|^4) + \ln(U_2|\eta(U)|^4)-\ln\kappa\Big]\cr
&\equiv\ 24\, \cFT(T,U)-24\, \ln\kappa\ ,\cr
}\eqn\Rtwo
$$
where $T\equiv B_{45}+i \sqrt{|G|}$ and $U\equiv (G_{45}+i
\sqrt{|G|})/G_{44}$  are the K\"ahler and complex structure moduli of
the two-torus, respectively. Since there are no contributions from the
$K3$ apart from the 24 zero modes, the result is proportional to the
topological partition function on $T^2$ \bcov. Indeed $\cFT(T,U)$ is
precisely what counts the 1/2 BPS states in the theory.

In the dual heterotic string on the six-torus $T^6$, the type IIA
modulus $T$ plays the r\^ole \doubref\md\onequarterBPS\ of the
heterotic dilaton: $T=\Sh$.  Thus \Rtwo\ represents a non-perturbative
result from the heterotic point of view, where the $\Sh$ dependence
reflects contributions from 1/2 BPS fivebrane instantons \harveymoore.

On the other hand, amplitudes sensitive to the intermediate, 1/4 BPS
states have not yet been computed, at least as far as we know.\foot{
Some works that deal with different but related issues discuss
topological amplitudes for six dimensional compactifications of type
IIA strings \BV, and the counting 1/8 BPS states of type IIA on $T^6$
\MMS. Moreover, counting 1/4 BPS states in 5d black holes has been
considered first in \StV.}   It is the purpose of the present paper to
compute certain of such couplings at one loop order for type IIA
strings on $K3\times T^2$, and investigate their structure.

More specifically, in the next section we will review some features of
1/4 BPS states in relation to the heterotic-type II duality.  In
section 3 we will then discuss (similar to refs.\
\doubref\elias\greekRtwo) some facts about BPS saturated helicity
traces and their relation to elliptic genera. In section 4 we will
first compute quartic couplings in the moduli fields that are sensitive
to the 1/2 BPS states only; not surprisingly, their structure turns out
to be essentially the same as for the $R^2$ coupling in \Rtwo.
Subsequently we will then compute certain sextic couplings, some of
which will be sensitive to 1/4 BPS states. We will find that these
couplings are characterized by two prepotentials $\cG_1(T,U)$ and
$\cG_2(T,\bar U)$, which enjoy an intriguing factorization property. In
their structure they resemble Borcherds-like sum formulas, with
counting functions given by the Eisenstein series $E_2(q)$. In the
subsequent sections we investigate these prepotentials by rewriting
them
in various ways, and also obtain a generalization to
an infinite sequence of $(2K\!+4)$-point amplitudes.

Finally, in the last section we will discuss the non-perturbative
significance of these amplitudes when they are mapped by duality to
the heterotic string. Specifically we will argue that the
prepotentials carry non-trivial information about genus $K$ instantons
on
heterotic fivebranes, and will also present some more speculative
remarks.

\chapter{Short and intermediate BPS multiplets}

We will consider \nex4 supersymmetric compactifications of type IIA
superstrings on $K3\times T^2$, or equivalently, of heterotic strings
on $T^6$. Besides the graviton, the bosonic content of the supergravity
 multiplet in N=4, $d=4$ supergravity is a complex scalar (the dilaton
$S_\het$) and six gravi-photons. In addition we have $22$ vector
multiplets, which contain each six scalars and one vector.  In terms of
the heterotic variables, the bosonic part of the action reads (up to
two derivatives, see e.g. \SS):
$$
\eqalign{S_{d=4,N=4}&\!=\!
\int d^4x \sqrt{-g}\lf[R+2 {\p^\mu S_\het\p_\mu \ov S_\het \o
(S_\het-\ov S_\het)^2}\ri.\cr
&\lf.\!-\!{1\o 4} \im(S_\het) F_{\mu\nu} LML F^{\mu\nu}\!\!+
{1\o 4} \re(S_\het) F_{\mu\nu} L \tilde F^{\mu\nu}+{1\o 8} \tr(\p_\mu
ML \p^\mu ML)\ri].\cr}
\eqn\action
$$
This action is manifestly invariant under $SO(22,6,\IR)$ \FKP, while
the equations of motion show a further invariance under $SL(2,\IR)$
acting on $S_\het$. Accordingly, the local geometry of the scalar
manifold is
$$
\cM\ =\ {SL(2,\IR) \o U(1)}\attac{S_\het} \!\!\!\!\!\!\!\times
{SO(22,6,\IR)\o SO(22,\IR)\times SO(6,\IR)}\ .
\eqn\modulispace
$$
The mass formula for 1/4 BPS states on this space is
\doubref\onequarterBPS\KK:
$$
\eqalign{M_{BPS}^2&=
{1 \o  S_\het-\ov S_\het} \Big[(\m+S_\het\n)^t (M+L)(\m+\ov S_\het
\n)\cr
&\pm\h \sqrt{[\m^t(M+L)\m][\n^t(M+L)\n]-[\m^t(M+L)\n]^2}\Big]\ ,\cr}
\eqn\BPS
$$
which involves the electric ($\m$) and magnetic ($\n$) charge vectors.
The sign is always meant to be chosen such that $M_{BPS}$ is maximized.
The degenerate case, in which the square root vanishes, corresponds to
the 1/2 BPS states \SS. Hence these may be viewed as specializations of
the more generic 1/4 BPS states that ``accidentally'' leave more
supersymmetries unbroken.

We will consider in the following only the subspace spanned by
$S_\het$, $T_\het$ and $U_\het$, so that the relevant moduli sub-space
is $\big({SL(2,\IR) \o U(1)}\big)^3$. On this subspace we have
$M_{BPS}^2\sim{1 \o {S_2}_\het {T_2}_\het{U_2}_\het} |Z|^2$, where
$Z=$max$\{|Z^+|,|Z^-|\}$ with \doubref\dieterBPS\KK:
$$
\eqalign{
Z^+&=m_1+m_2U_\het+k_1T_\het+k_2T_\het U_\het
\cr &\hskip 3cm +
 S_\het(n_1+n_2U_\het+p_1T_\het+p_2T_\het U_\het) \cr
Z^-&=m_1+m_2U_\het+k_1T_\het+k_2T_\het U_\het
\cr &\hskip 3cm +
\bar S_\het(n_1+n_2U_\het+p_1T_\het+p_2T_\het U_\het) \ .
}\eqn\inter
$$
One can check that in the degenerate case, $|Z^+|=|Z^-|$, these
central charges reduce to the 1/2 BPS mass formula \SS:
$$
Z=(q_1+q_2S_\het)(m_1+m_2U_\het+k_1T_\het+k_2T_\het U_\het)\ .
\eqn\short
$$
Note that the $S_\het$-independent terms of \inter\ and \short\
coincide, which implies that the perturbative states are at least 1/2
BPS and thus that the 1/4 BPS states are intrinsically
non-perturbative from the heterotic point of view. That is, there is no
perturbative calculation in the heterotic string that could possibly
see the 1/4 BPS states.

However, as noted above, the heterotic and type IIA compactifications
are dual to each other provided \refs{\md{,}\GPZ{,}\onequarterBPS}\ we
exchange: $T_\het=S$, $S_\het=T$ and $U_\het=U$. Inserting this into
\inter\ and \short, we see that at least some of the 1/4 BPS states do
have a perturbative description on the type IIA side, though certainly
not all of them.

Indeed we cannot expect to exactly compute the simultaneous dependence
on all three moduli $S,T,U$ in perturbation theory, in whatever
framework. But what we can do is to simply focus on the $T,U$
subspace on the type IIA side, while going to weak coupling, $S\equiv
T_\het\to i\infty$.  In this limit, further non-perturbative
corrections on the type IIA side  are suppressed. Even though this will
not capture the full story, it will capture at least some of the
non-perturbative physics in the heterotic string that is related to 1/4
BPS states.

The relevant physical states we thus consider are tensor  products of
the states on the $K3$ together with the momentum and winding modes on
$T^2$, characterized by
$$
\eqalign{
p_L&={1\o \sqrt{2T_2 U_2}}(m_1+m_2\bar U+n_1\bar T+n_2\bar T\bar U)\cr
p_R&={1\o \sqrt{2T_2 U_2}}(m_1+m_2\bar U+n_1 T+n_2 T\bar U)\ .
}\eqn\momenta
$$
These states are 1/4 BPS if either the left- or the right-moving
component is a ground state \refs{\KthreeBPS{,}\StV{,}\DVV},  ie.,
$N_L=h_L^\kt=0$ or $N_R=h_R^\kt=0$, where $N_{L,R}$ denotes the
oscillator number and $h_{L}^\kt+h_{R}^\kt$ the mass of excitations on
the $K3$. That is, (suppressing any vacuum energy shifts) the level
matching condition for BPS states reads
$$
|p_L|^2-|p_R|^2\ \equiv\ m_1n_2-n_1m_2\ =\
\cases{\ N_R+h_R^\kt\ \ : & 1/4 BPS \cr -N_L-h_L^\kt\ :
& 1/4 BPS \cr\ 0
\ \ \ \ \qquad\qquad: & 1/2 BPS}\ ,
\eqn\BPSLM
$$
which exhibits the dyonic nature of the 1/4 BPS states.
Clearly, the 1/2 BPS states correspond to both left and right moving
ground states, and the level matching condition is identically
satisfied for the momenta \short\ with $k_i=0$.

\chapter{Topological indices and \nex4 supersymmetry}

One--loop amplitudes that are sensitive only to BPS states must
certainly very special. Indeed they must be proportional to certain
``helicity traces'' \refs{\HM{--}\GV}\ (or generalizations thereof), in
which the long multiplets cancel out. A canonical example for such
a trace is given by \elias:
$$
\langle\,\l^{2n}\,\rangle\ =\
\Big({\del\over\del v_L}+{\del\over\del \bar v_R}\Big)^{2n}
\mathop{{\rm Str}}_{{\rm {all\hfil\atop sectors}}}
\Big[q^{L_0}\bar q^{\bar L_0}e^{v_L J_{L}\st}e^{\bar v_R J_{R}\st}
\Big]\attac{v_L=\bar v_R=0}.
\eqn\heltrace
$$
Here, $\l\equiv J_{L}\st+J_{R}\st$ denotes the helicity operator, where
in each left and right moving sector $J\st={1\over 2\pi i}\oint
\tilde\psi^{\bar\mu}\psi^\mu$ is the zero mode of the fermion number
current (``(st)'' denotes the light-cone space-time part  of the
theory).

In order to recognize the saturation or vanishing of such traces more
easily, it is convenient to map them to the RR sector of the theory, in
which this becomes a simple question of saturation of fermionic zero
modes. This map \doubref\WL\ghosttrial\  is universal for a given
number of left- and right-moving supersymmetries and otherwise does not
depend on the background. For simplicity, we will write the relevant
identity down only for the left-moving variables, understanding that an
analogous identity holds independently also in the right-moving sector:
$$
\eqalign{
\mathop{{\rm Str}}_{{\rm {all\atop sectors}}}
&\Big[q^{L_0}e^{v\st J\st +v^{(T2)}J^{(T2)}+v^{\kt}J^{\kt}}\Big]
\cr\ &=\
\mathop{{\rm Tr}}_{{\rm {R}}}
\Big[(-1)^{J\st+J^{(T2)}+J^{\kt}}q^{L_0}
e^{\hat v\st J\st +\hat v^{(T2)}J^{(T2)}+\hat v^{\kt}J^{\kt}}
\Big]\ ,
}\eqn\RI
$$
where
$$
\eqalign{
\hat v\st\ &=\ \shalf v\st+\shalf v^{(T2)}+\shalf\sqrt2 v^{\kt}\cr
\hat v^{(T2)}\ &=\ \shalf v\st+ \shalf v^{(T2)}- \shalf\sqrt2
v^{\kt}\cr
\hat v^{\kt}\ &=\  \shalf \sqrt2 v\st- \shalf \sqrt2 v^{(T2)}\ .
}\eqn\trialitymap
$$
Above, $J^{(T2)}$ denotes the fermion number current in the $T^2$
sector and $J^{\kt}$ the zero mode of the $U(1)\subset SU(2)$ current
of the \nex4 world-sheet superconformal algebra that is intrinsic to a
sigma-model on $K3$.
{}From this it is easy to see that in the type IIA compactification on
$K3\times T^2$, we need at least two current insertions in each of the
left and right moving sectors in order to get a non-vanishing result.

There are however traces that are more general than
the left-right symmetric helicity trace in \heltrace, and involve
arbitrarily high powers of a current insertion in one of the
left- or right-moving sectors, for example:
$$
\eqalign{
B_{2K+4}\ &\equiv\
\big\langle\,(J_L\st)^{2+2K}(J_R\st)^{2}\,\big\rangle\ =\
\Big({\del\over\del v_L}\Big)^{2+2K}\Big({\del\over\del \bar
v_R}\Big)^{2}\ \times
\cr
\mathop{{\rm Tr}}_{{\rm {RR}}}\!\Big[
(-1&)^{\sum (J_L^i+J_R^i)}
q^{L_0}\bar q^{\bar L_0}
e^{{1\over2}     v_L(J_{L}\st+J_L^{(T2)}+\sqrt2 J_L^{\kt})}
e^{{1\over2}\bar v_R(J_{R}\st+J_R^{(T2)}+\sqrt2 J_R^{\kt})}
\Big]\attac{v_L=\bar v_R=0} \cr
&= \
|\eta(q)|^{-12}Z_{2,2}(T,U,q,\bar q)
\Big({\del\over\del v_L}\Big)^{2+2K}\Big({\del\over\del \bar
v_R}\Big)^{2}
\theta_1(\shalf v_L)^2 \bar\theta_1(\shalf \bar v_R)^2\ \times
\cr &\ \ \ \ \
\mathop{{\rm Tr}}_{{\rm {RR}}}
\Big[
(-1)^{J_L^{\kt}+J_R^{\kt}}
q^{L_0}\bar q^{\bar L_0}
e^{{1\over2}\sqrt2 (v_L J_L^{\kt}+\bar  v_R J_R^{\kt})}
\Big]\attac{v_L=\bar v_R=0} \cr
&=\
2\eta(q)^{-6}Z_{2,2}(T,U,q,\bar q)
\Big({\del\over\del  v_L}\Big)^{2+2K}\!\!
\CHI_{({\rm st}\times T^2)}(\shalf v_L,q)
\CHI_{\kt}(\shalf v_L,q)\attac{v_L=0}\ .
}\eqn\thebigtrace
$$
Here,
$$
Z_{2,2}(T,U,q,\bar q)\ =\ \sum_{p_L,p_R} q^{{1\over2} |p_L|^2}
\bar q^{{1\over2}|p_R|^2}
\eqn\Ztwotwo
$$
is the partition function of windings and momenta on the two-torus, and
$$
\eqalign{
\CHI_{({\rm st}\times T^2)}(v,q)\ &=\ \ \Big(\!{i\theta_1(v,q)\over
\eta^3(q)}\Big)^2\ \ \
=:\ \sum_{n\geq0,\ell\in\ZZ}d(4n-\ell^2)\,q^n e^{\ell v}\cr
\CHI_{\kt}(v,q)\ &=8 \sum_{i=2}^4
\Big({\theta_i(v,q)\over \theta_i(0,q)}\Big)^2
\ =:\ \sum_{n\geq0,\ell\in\ZZ}e(4n-\ell^2)\,q^n e^{\ell v}\cr
}\eqn\ellgen
$$
are the elliptic genera \ellg\ of the space-time sector times $T^2$ and
of the $K3$ surface, respectively.  Like all elliptic genera, these are
weak Jacobi modular forms, namely of weight $-2$ and index $1$, and of
weight $0$ and index $1$, respectively.

Even more general traces can be obtained by inserting the individual
fermion number currents $J_{L}\st$, $J_L^{(T2)}$ and $J_L^{\kt}$
independently. The generalized Riemanm identity \RI\ will then in
general produce some product of the individual elliptic genera
$\CHI_{({\rm st})}(\hat v{\st},q)$, $\CHI_{(T^2)}(\hat v^{(T^2)},q)$
and
$\CHI_{\kt}(\hat v^{\kt},q)$. As we will see, for the amplitudes that
we will consider, only one of those factors will be realized.

Elliptic genera depend
holomorphically on $q$, which reflects that all non-zero modes in the
right-moving sector cancel out due to world-sheet supersymmetry.
Via the identity \RI\ (which is due to space-time supersymmetry), this
is simultaneously a reflection of the fact that the long multiplets
cancel out in the trace, independent of any deformations in
the $K3$ moduli.  While in fact the total number of BPS states may jump
when varying the moduli, the weighted helicity sums count effectively
net numbers of BPS states and so remain invariant. It is this
index-like, topological nature of BPS saturated amplitudes what makes
them special and their modular integrals exactly computable
\refs{\WL,\HM{--}\elias}.

More specifically, for $K=0$ only the $\chi(K3)=24$ left and right
moving ground states can contribute in the $K3$ sector, so that
$$
\eqalign{
B_4\ =\ \big\langle\,(J_L\st)^2(J_R\st)^{2}\,\big\rangle\
&=\ 4 \ Z_{2,2}(T,U,q,\bar q) \ \
\mathop{{\rm Tr}}_{K3}\Big[(-1)^{J_L^{\kt}+J_R^{\kt}}
q^{L_0}\bar q^{\bar L_0}\Big]\cr
&=\ 96\ Z_{2,2}(T,U,q,\bar q)
}\eqn\bfourtrace
$$
gets only contributions from 1/2 BPS states and this is what underlies
the $R^2$ coupling \Rtwo\ \doubref\harveymoore\greekRtwo.

On the other hand, for $K>0$ in \thebigtrace\ the states that
contribute to the trace consist of right-moving ground states and
arbitrary left-moving states -- from what we said in the previous
section, this precisely characterizes the 1/4 BPS states.  For example,
for $K=1$ one has the following six-th order trace:\foot{Which has
previously been calculated in the ${\bf Z}_2$ orbifold limit of $K3$
\doubref\greekRtwo\elias.}:
$$
\eqalign{
B_6\ &=\ \big\langle\,(J_L\st)^4(J_R\st)^{2}\big\rangle
\cr &=
12\,  Z_{2,2}(T,U,q,\bar q)
\ \ \mathop{{\rm Tr}}_{K3}\Big[(-1)^{J_L^{\kt}+J_R^{\kt}}
(J_L^{\kt})^2
q^{L_0}\bar q^{\bar L_0}\Big]
\cr &=
192\,  E_2(q)\,Z_{2,2}(T,U,q,\bar q)\ .
}\eqn\bsixtrace
$$

The issue is now to identify physical amplitudes that contain
these building blocks.

\chapter{$1/2$ and $1/4$ BPS saturated amplitudes }

As a warm--up, we study quartic interactions of the
$T$ and $U$ moduli at one--loop order in type IIA compactified on
$K3\times T^2$:
$$
\vev{V_{\phi_1}(k_1)V_{\phi_2}(k_2) V_{\phi_3}(k_3)V_{\phi_4}(k_4)}
\ ,\ \quad\phi_i=T,U,\ov T,\ov U\ .
\eqn\amplitudei
$$
We want to extract from \amplitudei\  the kinematical factor
$(k_1k_3)(k_2k_4)$ (and permutations thereof),  which corresponds to an
one--loop corrections $\cA$ to the term $(\p_\mu\phi_1\p^\mu\phi_3)
(\p_\nu\phi_2\p^\nu\phi_4)$  in the effective action. These amplitudes
receive non--vanishing contributions only from the $NSNS$--sector and
not from the $RNS,NSR$  and $RR$--sectors. We thus consider insertions
of the moduli vertex operators:
$$
\eqalign{
V^{(0,0)}_T(k)&={2\o T-\ov T}:[\ov\p \ov Z+
i(k\tilde\psi)\ov{\tilde\Psi}(\ov z)]\
[\p Z+i(k\psi)\Psi(z)]\ e^{ik_\mu X^\mu(z,\ov z)}:\ ,\cr
V^{(0,0)}_U(k)&={-2 \o U-\ov U}:[\ov\p
Z+i(k\tilde\psi)\tilde\Psi(\ov z)]\
[\p Z+i(k\psi)\Psi(z)]\ e^{ik_\mu X^\mu(z,\ov z)}:\ ,\cr}
\eqn\vertexnsns
$$
in the zero ghost picture with $\ov Z=\sqrt{T_2/2U_2}(X^4+UX^5), \
Z=\sqrt{T_2/2U_2}(X^4+\ov U X^5),\
\ov\Psi=\sqrt{T_2/2U_2}(\psi^4+U\psi^5),\
\Psi=\sqrt{T_2/2U_2}(\psi^4+\ov U\psi^5)$. In this normalization we
have: $\vev{\Psi(z_1)\ov\Psi(z_2)}_{\al,even}=
-{\th_\al(z_{12},\tau)\th_1'(0,\tau)\o
\th_\a(0,\tau)\th_1(z_{12},\tau)}$,
$\vev{\Psi(z_1)\Psi(z_2)}_{\al,even}=0$, and for the kinematics we
consider,
the only non--vanishing fermionic contractions are those that lead to
the helicity trace $B_4$ in \bfourtrace.  However, from the bosonic
contractions\foot{The third correlator may also contain
a $\delta$--function \minahan. However, it would lead to manifestly
non--covariant amplitudes.}:
$$
\eqalign{
\vev{\ov\p Z(\ov z_1,z_1)\ \ov\p Z(\ov z_2,z_2)}&=p_R^2\cr
\vev{\ov\p Z(\ov z_1,z_1)\ \p Z(\ov z_2,z_2)}&=p_Rp_L\cr
\vev{\ov\p Z(\ov z_1,z_1)\ \p \ov Z(\ov z_2,z_2)}&=p_R\ov p_L\cr
\vev{\ov\p Z(\ov z_1,z_1)\ \ov \p \ov Z(\ov z_2,z_2)}&=|p_R|^2-
{1\o 2\pi\tau_2}+{1\o2\pi^2}\p^2_{\ov z_{21}}G_B\cr}
\eqn\vevs
$$
($G_B\equiv-\ln|\chi|^2$), we have additional Narain momentum
insertions, whose contributions are crucial for modular invariance. The
resulting modular  integrals\foot{Some of these integrals have to be
regularized with an IR--regulator. This is described in the appendix
of \FS\ and results in an extra constant contribution.} can be
evaluated by using extensively the results of \FS.  A typical example
for a non--vanishing amplitude is:
$$
\eqalign{
U_2^4\ {\cA}_{(\p_\mu U\p^\mu\ov U)(\p_\nu U\p^\nu\ov U)}&\ =
T_2^2U_2^2\ {\cA}_{(\p_\mu T\p^\mu\ov T)(\p_\nu U\p^\nu\ov U)}\cr
&=\!\!\int\!{d^2\tau \o\tau_2}\sum_{(p_L,p_R)}\!\!\!\tau_2^2
\lf(|p_R|^2\!-\!{1\o 2\pi \tau_2}\ri)\lf(|p_L|^2\!-\!{1\o 2\pi
\tau_2}\ri)
q^{\h |p_L|^2} \ov q^{\h |p_R|^2}
\cr
&=-{1\o 4\pi^2}\,\cFT(T,U)+{1\o 4\pi^2}\, [1+\gamma_E-\ln(4\pi)]\ .}
\eqn\aone
$$
More generally,  we find that all the non-vanishing amplitudes are
proportional to $\cFT(T,U)$ (cf., \Rtwo), which of course reflects
that the helicity trace $B_4$ is sensitive only to 1/2 BPS states.

Now let us turn to more interesting scalar field interactions at the
sixth derivative level. More specifically, we consider the
following type of amplitudes:
$$
\vev{V_{\phi_1}(k_1)V_{\phi_2}(k_2)V_{\phi_3}(k_3)V_{\phi_4}(k_4)
V_{\phi_5}(k_5)V_{\phi_6}(k_6)}\ ,\ \quad\phi_i=T,U,\ov T,\ov U
\eqn\amplitudeii
$$
and focus on the kinematics for which each modulus $\phi_i$ contributes
one  momentum $k_i$.  This momentum structure can arise in three
different ways: $(i)$~four--fermionic contractions on both sides,
giving rise to the helicity trace $B_4$,  $(ii)$ eight-- and
four--fermionic contractions on the right- and left-moving sides,
respectively, or $(iii)$ four-- and eight--fermionic  contractions on
the right-moving and left-moving sides, respectively.

\goodbreak

The main technical issue is the evaluation of the modular integrals,
which is not entirely trivial and will be outlined in Appendix \appA.
Performing these integrals, it turns out that there are two types of
non--vanishing results, one type displaying again only 1/2 BPS states,
the other however being sensitive to 1/4 BPS states. As an example for
the first type, consider
$$
\eqalign{
U_2^6\cA^{(i)}_{(\p_\mu U\p^\mu \ov U)(\p_\nu U\p^\nu\ov U)
(\p_\rho \phi\p^\rho\ov \phi)}&=
-\int{d^2\tau \o\tau_2}\sum_{(p_L,p_R)}
\tau_2^4\lf(|p_L|^4-{2\o\pi\tau_2}|p_L|^2+{1\o 2\pi^2\tau_2^2}\ri)\cr
&\hskip 20pt\times\lf(|p_R|^4-{2\o\pi\tau_2}|p_R|^2+
{1\o 2\pi^2\tau_2^2}\ri)\ q^{\h |p_L|^2} \ov q^{\h |p_R|^2}\cr
&={1\o 4\pi^4}\,\cFT(T,U)-{1\o 8\pi^4}\,[3+2\gamma_E-2\ln(4\pi)]\ ,\cr}
\eqn\basiceight
$$
where $\phi$ can be any of $\{T,U\}$.  Since it involves the eight
fermion contraction of type $(i)$, which gives rise to $B_4$, this
amplitude is obviously sensitive only to 1/2 BPS states. However, it
can happen even for amplitudes with twelve-fermion contractions of
types $(ii)$ and $(iii)$ that only 1/2 BPS states contribute. An
example is given by the following two contributions to the same
amplitude\foot{The normalization of the vertex operators \vertexnsns\
is absorbed into $\DD$.}:
$$
\eqalign{
\DD^{(ii)}_{(\p_\mu U\p^\mu \ov U)(\p_\nu U\p^\nu\ov U)
(\p_\rho \phi\p^\rho\ov \phi)}&=
\int{d^2\tau \o\tau_2}\sum_{(p_L,p_R)}\tau_2^4
\lf(|p_R|^2-{1\o 2\pi\tau_2}\ri)
\cr
&\hskip 20pt\times\lf(|p_L|^4-{2\o\pi\tau_2}|p_L|^2+{1\o
2\pi^2\tau_2^2}\ri)
\hat{\bar E_2}\ q^{\h |p_L|^2} \ov q^{\h |p_R|^2}\cr
&=-{3\o 4\pi^4}\,\cFT(T,U)+{3\o 8\pi^4}\,[3+2\gamma_E-2\ln(4\pi)]\ ,\cr
\DD^{(iii)}_{(\p_\mu U\p^\mu \ov U)(\p_\nu U\p^\nu\ov U)
(\p_\rho \phi\p^\rho\ov \phi)}&=
\int{d^2\tau \o\tau_2}\sum_{(p_L,p_R)}\tau_2^4
\lf(|p_R|^4-{2\o\pi\tau_2}|p_R|^2+{1\o 2\pi^2\tau_2^2}\ri)\cr
&\hskip 20pt\times\lf(|p_L|^2-{1\o 2\pi\tau_2}\ri)\E_2\
q^{\h |p_L|^2} \ov q^{\h |p_R|^2}\cr
&=-{3\o 4\pi^4}\,\cFT(T,U)+{3\o 8\pi^4}\,[3+2\gamma_E-2\ln(4\pi)]\ .}
\eqn\beforecargese
$$
Even though the (suitable regularized)
second derivative of the elliptic genus appears in the
form of $\E_2\equiv E_2-{3\o \pi \tau_2}$, the integral involving $E_2$
vanishes\foot{Up to a term ${T_2\o 4\pi^3} c(0)$, which is absorbed in
$\ln(\eta(T)$).} in a non-trivial manner, so that basically only the
non--harmonic part of $\E_2$ contributes -- this means that again
only 1/2 BPS and no 1/4 BPS states contribute.

Summarizing, the first kind of sextic couplings has exactly the same
1/2 BPS structure as the quartic couplings discussed above.\foot{It is
known \GKPR\ that some of the six-derivative couplings are related to
the four-derivative ones by field redefinitions. The similarity of the
results in \beforecargese\  and \aone\ may be partly related to that.
Moreover, their kinematical structure coincides with some of the
six-derivative couplings that arise from expansions of Born-Infeld
actions~\MKPCR, and so may be also reproduced from simple $D$-brane
interactions.}

On the other hand, in the following examples, where only one type of
contraction contributes (either of type $(ii)$ or type $(iii)$), we see
an interesting new structure emerging. More specifically, we
find\foot{However: $\DD^{(i)}_{(\p_\mu T\p^\mu U)(\p_\nu T\p^\nu U)
(\p_\rho \phi\p^\rho\ov \phi)}=0=\DD^{(i)}_{(\p_\mu T\p^\mu U)(\p_\nu
T\p^\nu U)
(\p_\rho T\p^\rho\ov U)}$.}:
$$
\eqalign{
\DD^{(ii)}_{(\p_\mu T\p^\mu U)(\p_\nu T\p^\nu U)
(\p_\rho \phi\p^\rho\ov \phi)}\hskip-50pt&\cr
&=\int{d^2\tau \o\tau_2}\sum_{(p_L,p_R)}\tau_2^4
\lf(|p_R|^2-{1\o 2\pi\tau_2}\ri)p_L^4\ \hat{\bar E}_2
\ q^{\h |p_L|^2} \ov q^{\h |p_R|^2}\cr
&=-{4\o\pi^2}T_2^2U_2^2\
\lf(\p_{U}+{2\o U-\ov U}\ri)\lf(\p_{T}+{2\o T-\ov T}\ri)\cG_1(T,U)\cr
\DD^{(ii)}_{(\p_\mu T\p^\mu U)(\p_\nu T\p^\nu U)
(\p_\rho T\p^\rho\ov U)}\hskip-50pt&\cr
&=\int{d^2\tau \o\tau_2}\sum_{(p_L,p_R)}\tau_2^4
\ov p_R^2p_L^4\ \hat{\bar E}_2
\ q^{\h |p_L|^2} \ov q^{\h |p_R|^2}\cr
&={4\o\pi^2}T_2^3U_2\
\lf(\p_{T}+{4\o T-\ov T}\ri)\lf(\p_{T}+{2\o T-\ov T}\ri)\cG_1(T,U)\ ,}
\eqn\athreeHol
$$
and:
$$
\eqalign{
\DD^{(iii)}_{(\p_\mu T\p^\mu \ov U)(\p_\nu T\p^\nu\ov U)
(\p_\rho \phi\p^\rho\ov \phi)}\hskip-50pt&\cr
&=\int{d^2\tau \o\tau_2}\sum_{(p_L,p_R)}\tau_2^4
\lf(|p_L|^2-{1\o 2\pi\tau_2}\ri)\ov p_R^4\ \E_2
\ q^{\h |p_L|^2} \ov q^{\h |p_R|^2}\cr
&=-{4\o\pi^2}T_2^2U_2^2\
\lf(\p_{\ov U}-{2\o U-\ov U}\ri)\lf(\p_{T}+{2\o T-\ov T}\ri)\cG_2(T,\ov
U)\cr
\DD^{(iii)}_{(\p_\mu T\p^\mu \ov U)(\p_\nu T\p^\nu\ov U)
(\p_\rho T\p^\rho U)}\hskip-50pt&\cr
&=\int{d^2\tau \o\tau_2}\sum_{(p_L,p_R)}\tau_2^4
p_L^2\ov p_R^4\ \E_2\ q^{\h |p_L|^2} \ov q^{\h |p_R|^2}\cr
&={4\o\pi^2}T_2^3U_2\
\lf(\p_T+{4\o T-\ov T}\ri)\lf(\p_{T}+{2\o T-\ov T}\ri)\cG_2(T,\ov U)\
.}
\eqn\athreeAntiHol
$$
Furthermore\foot{But,\ \
$\DD^{(i)}_{(\p_\mu T\p^\mu \ov T)(\p_\nu T\p^\nu\ov U)
(\p_\rho \phi\p^\rho\ov \phi)}=
0$.}
$$
\eqalign{
\DD^{(iii)}_{(\p_\mu T\p^\mu \ov T)(\p_\nu T\p^\nu\ov U)
(\p_\rho \phi\p^\rho\ov \phi)}&=
\int{d^2\tau \o\tau_2}\sum_{(p_L,p_R)}\tau_2^4
\lf(|p_L|^2-{1\o 2\pi\tau_2}\ri)\lf(|p_R|^2-{3\o 2\pi\tau_2}\ri)\cr
&\hskip 20pt\times \ov p_R^2\ \E_2\ q^{\h |p_L|^2} \ov q^{\h
|p_R|^2}\cr
&=-{1\o\pi^2} T_2U_2\ \cG_2(T,\ov U)\ .\cr}
\eqn\additional
$$
One can easily check that there is no tree-level contribution to this
kind of BPS saturated amplitudes, so they are exact up to one loop
order (possible higher loop and non-perturbative corrections are
suppressed anyway in the limit that we consider, $S\to i\infty$).

Importantly, what characterizes these amplitudes is a prepotential:
$$
\eqalign{
\cG_1(T,U)\ &=\
{\zeta(-1)\o 2} c(0)+
{i\o 4\pi} {c(0)\o U-\ov U}+{i\o 4\pi} {1\o T-\ov T}
+{3\o 2\pi^2}{1\o (T-\ov T)(U-\ov U)}\cr
&-{3\o\pi^2(T-\ov T)}\ \p_{U}\sum_{l>0}\
\Li_1\lf(e^{2\pi ilU}\ri)
-{3\o\pi^2(U-\ov U)}\ \p_T\sum_{k>0}\ \Li_1\lf(e^{2\pi ikT}\ri)
\cr
&+\sum_{(k,l)>0}c(kl)\ \Li_{-1}\lf(e^{2\pi i(kT+lU)}\ri)
\ ,}
\eqn\prep
$$
and similarly for $\cG_2(T,\bar U)$ in the chamber $T_2>U_2$. Here
$\zeta(-1)=-{1\o 12}$, and the poly-logarithms are defined by
$\Li_a(z)=\sum_{p>0}z^pp^{-a}$ for $a>0$ and
$\Li_a(z)=(z\del_z)^{|a|}{1\over1-z}$ for $a<0$; in particular,
$\Li_1(e^z)=-\ln(1-e^z),\ \Li_0(e^z)={e^z\o 1-e^z}$ and
$\Li_{-1}(e^z)={e^z\over(1-e^z)^2}$. Moreover the sum runs over the
positive roots $k>0,\ l\in \ZZ\ \ \wedge\ \ k=0,\ l>0$, and the
coefficients are defined by:
$$
\sum_n c(n)q^n\ :=
\ E_2(q)\ \equiv\ 1-24q-72q^2+\ldots \ .
\eqn\counting
$$
This must be the derivative of some combination of the elliptic genera
in \ellgen, but because of the uniqueness of the quasi-modular form of
weight two, it is unclear at this point of exactly which elliptic
genus:\foot{It was the equality of these expressions that has been
misleading us to some conclusions that we have presented in a previous
version of this paper.} $E_2=\coeff14{\del_v}^2
\CHI_{\kt}(v/2,q)|_{v=0}= \coeff12{\del_v}^4 \CHI_{({\rm st}\times
T^2)}(v/2,q)|_{v=0}= \coeff1{96}{\del_v}^4 \CHI_{({\rm st}\times
T^2)}(v/2,q)\times$ $ \CHI_{\kt}(v/2,q)|_{v=0}$. While the distinction is not
important here, it will be more relevant later on when we will discuss
the generalization to $(2K+4)$-point amplitudes.

Note that since $c(-1)=0$, there is no singularity in the $T,U$ moduli
space and this reflects the impossibility of states becoming massless.
Note also that $\cG_1(T,U)$ has weight $2$ under $T$--and $U$--duality,
respectively, and indeed we find that \prep\ and its
holomorphic/anti-holomorphic cousin can be rewritten in terms of a
simple product involving (regularized) Eisenstein functions:
$$
\eqalign{
\cG_1(T,U)\ &=\  - \Coeff1{24}\E_2(T)\E_2(U)\cr
\cG_2(T,\ov U)\ &=\  - \Coeff1{24}\E_2(T)\E_2(\ov U)\ .
}\eqn\productform
$$
These intriguing identities exhibit a factorization that
is not manifest in \prep. We can furthermore obtain both
of these prepotentials (and their complex conjugates) by taking mixed
derivatives of the following function:
$$
{\cH}(T,U)\ =\ -6\,\ln(T_2|\eta(T)|^4) \cdot \ln(U_2|\eta(U)|^4)\ ,
\eqn\GTU
$$
which in this sense appears to be a more fundamental
function\foot{ Switching on all the moduli should promote
$\ln(U_2|\eta(U)|^4)$ to the logarithm of some $SO(22,6,\ZZ)$ modular
form.} for the six-point amplitudes we consider here. It is the analog
of the 1/2 BPS free energy $\cFT(T,U)$ in \Rtwo, the difference being
that the $\ln\eta$'s are multiplied rather than added; by adjusting
possible integration constants we see that ${\cH}(T,U)$ is essentially
the square of $\cFT(T,U)$.

\chapter{Partition functions}

The holomorphic prepotential \prep\ is one of the main results of this
paper. Its appearance hints at the existence of a yet unknown
superspace formulation of the theory, in which it might figure as an
effective lagrangian. It resembles the ``Borcherds'' type prepotentials
that arise in other contexts \refs{\HM{,}\kawai{,}\Gm{,}\LS}, where
non-negative poly-logarithms appear instead. However, that difference
is not important and simply to be attributed to the mass dimension of
the couplings.

The structurally more profound feature is that $\cG_1(T,U)$
intrinsically mixes the K\"ahler and complex structure sectors, and
this is specifically tied to the 1/4 BPS states.  Indeed, when
restricted to the subset of 1/2 BPS states (which corresponds to the
terms with $kl=0$), the sum in \prep\ nicely separates into decoupled
pieces:
$$
\eqalign{
\cG_1(T,U)\ \ \
&\mathop{{\longrightarrow}}^{{1/2 BPS}}_{(kl)=0}\ \ \
\sum_{k=1}^\infty
{ e^{2\pi i kT} \over (1-e^{2\pi i kT})^2}
+\sum_{l=1}^\infty
{ e^{2\pi i lU} \over (1-e^{2\pi i lU})^2}+\dots
\cr
&=\ -{1\over 2\pi i}\big[\del_T \ln\eta(T)+\del_{U}
\ln\eta(U)\big]+\dots\cr
&=\ -{1\over 4\pi i}\big(\del_T+\del_{U}\big)\,\cFT(T,U)
+\dots\ ,
}\eqn\onehalfsplit
$$
which in turn can be written manifestly in terms
of the 1/2 BPS spectrum using \doubref\fklz\coniA:
$$
\cFT(T,U)\ =\
-\coeff1{12}\!\!\!\!\sum_{m_1n_2-n_1m_2=0} \ln|p_R|^2=
-\coeff1{12}\!\!\!\!
\sum_{m_1n_2-n_1m_2=0} \ln|p_L|^2\ .
\eqn\onehalfBPSsum
$$
We thus explicitly see  that the mixing terms in $\cG_1(T,U)$ (or
$\cG_2(T,\bar U)$) correspond to the 1/4 BPS states and originate from
the presence of $E_2(\bar q)$ (or $E_2(q)$) in the integrand.  Its
effect is to shift the 1/2 BPS level matching condition,
$|p_L|^2=|p_R|^2$, to the 1/4 BPS condition: $|p_L|^2=|p_R|^2+kl$ (or
$|p_R|^2=|p_L|^2+kl$).

Using the product representation \productform\ and the well-known sum
formulas of the Eisenstein series, we can represent the prepotentials
in a form that generalizes the 1/2 BPS sum \onehalfBPSsum:
$$
\eqalign{
-24\,\cG_1(T,U)\attac{{holom.\atop piece}}
\!\!\!\!\!\!=\ E_2(T)E_2(U)\
&=\! \sum_{{(N_1,N_2)\neq(0,0)\atop (M_1,M_2)\neq(0,0)}}
           {1\o (N_2+N_1T)^2(M_2+M_1 U)^2}\cr
&= {1\o 2T_2U_2}\!\sum_{m_1n_2-n_1m_2=0}
{\c(m_i,n_i)\o \ov p_L^2}\ ,\cr
-24\,\cG_2(T,\ov U)\attac{{holom.\atop piece}}
\!\!\!\!\!\!=\ E_2(T)E_2(\ov U)\ &=
{1\o 2T_2U_2}\!\!\sum_{m_1n_2-n_1m_2=0} {\c(m_i,n_i)\o p_R^2}\ ,}
\eqn\bpssumii
$$
with $m_1=N_2M_2, m_2=N_2M_1, n_1=N_1M_2, n_2=N_1M_1$. Since there are
in general many different $\{M_i,N_i\}$ that contribute to a given set
$\{m_i,n_i\}$, the coefficients $\c(m_i,n_i)$ are in general larger
than one, and this must be so since otherwise the sums would be
counting (just like \onehalfBPSsum) exactly the 1/2 BPS states.

Note that while $\cFT(T,U)$ has been written in \onehalfBPSsum\ as a
sum over 1/2 BPS states circulating in loops, it has also an
interpretation in terms of world-sheet instantons \bcov; this is
exhibited by the instanton expansion in the first line of eq.\
\onehalfsplit.  Such a view-point is indeed more natural in the path
integral formulation, where $\cFT(T,U)$ is seen as counting holomorphic
maps from a toroidal world-sheet into the target space~$T^2$.

The additional mixing terms proportional to $e^{2\pi i kT}e^{2\pi i
lU}$, which are due to the 1/4 BPS states, must have an analogous
instantonic interpretation,  however involving holomorphic (and
anti-holomorphic) maps that couple together both K\"ahler and complex
structure sectors. Such configurations can presumably be obtained via
$T$-duality from string networks  \doubref\johnS\network, in which the
1/4 BPS states have a simple geometric representation.\foot{By
correctly identifying the variables, we can map the mass formula for
(wrapped) triple string junctions to the mass formula \momenta, ie.,
$M^2_{BPS}=\sum_{i=1}^3T_{p_i,q_i}={\rm max}\{|Z^+|^2,|Z^-|^2\}$. Here,
$p_i,q_i$ are the charges of the $i$-th link of the junction,
$T_{p_i,q_i}$ the corresponding tension and $Z^\pm$ the central charges
of \inter. Our results thus amount to counting such string junctions.}

\chapter{Generalization to an infinite sequence of prepotentials}

We will now discuss the generalization of the results of  section 4 to
$(2K\!+4)$--point amplitudes with $K+1$ pairs of $T$ and $U$ moduli,
besides one pair of moduli, $(\phi,\ov\phi)$ with $\phi=\{T,U\}$.
Performing four-fermion contractions  in the left-moving sector and
$4K\!+4$ fermion contractions in the right-moving sector, integrating
over the location of the vertex operators and subsequently applying the
Riemann identity (see Appendix \appB\ for some of the details), we
eventually find:
$$
\eqalign{
\tilde{\cA}_{(\p_\nu T\p^\nu U)^{K+1}
(\p_\rho \phi\p^\rho\ov \phi)}^\fk \ &=
\ {1\o 4}\int{d^2\tau\o \tau_2}\sum_{(p_L,p_R)}
\lf(|p_R|^2-{1\o 2\pi\tau_2}\ri)
(\tau_2p_L)^{2K+2} q^{\h |p_L|^2} \ov q^{\h |p_R|^2}
\ \cr
&\qquad \qquad\times\ \Big({\del\over\del \bar  v}\Big)^{2K+2}
\lf[e^{-{\bar v^2\o 4\pi\tau_2}}
\CHI_{({\rm st}\times T^2)}(\bar v,\bar q)\ri]\lf.\ri|_{\bar v=0}\ .
}\eqn\higher
$$
The most harmonic part then evaluates to:
$$
-{1\o \pi^2}(-2T_2U_2)^{K+1}
\lf(\p_T+{2K\o T-\ov T}\ri)\lf(\p_U+{2K\o U-\ov U}\ri)\cG_1^\fk(T,U)\ ,
\eqn\harmA
$$
with prepotentials $\cG_1^\fk(T,U)$ that form an infinity
sequence given by:
$$
\cG_1^\fk(T,U)
\attac{{harm.\atop piece}}\!\!\!\!\!\!\!\!\!\!
\ =\
\Coeff12\zeta(1\!-\!2K)\,c^\fk(0)
+\!\!\sum_{(k,l)>0}c^\fk(kl)\ \Li_{1-2K}\lf(e^{2\pi i(kT+lU)}\ri).
\eqn\moregeneralK
$$
The counting functions for these are simply:
$$
\sum_n c^\fk(n)q^n\ =\
\Big({\del\over\del v}\Big)^{2K+2}\!
\CHI_{({\rm st}\times T^2)}(v,q)\attac{v=0}\!\!\!\!\!\! =\
\Big({\del\over\del v}\Big)^{2K+2}\!
\!\left(\!{i\theta_1(v,q)\over\eta^3(q)}\right)^2
\!\attac{v=0}\!\!\!\!\!,
\eqn\delE
$$
In fact, we can concisely assemble all the prepotentials into a single
generating function:
$$
\eqalign{
\widehat\cG_1(T,U,V)\ &=\
\sum_{K=0}^\infty {1\over(2K)!} V^{2K} \cG_1^\fk(T,U)\cr
&=\ \sum_{(k,l,m)>0}d(4kl-m^2)\ \Li_{3}\lf(e^{2\pi i(kT+lU)}e^{mV}\ri)\
,
}\eqn\thefullthing
$$
where $d(4n-m^2)$ are the expansion coefficients \ellgen\ of
$\cE_{({\rm st})\times T^2}(v,q)$.

Note that what appears here is the elliptic genus of the space-time
sector times $T^2$, and not the elliptic genus of $K3$ as one might
have expected.\foot{and as we had mis-stated in a previous version of
this paper. It does however appear in subsequent work \morales\ which
deals with graviphoton amplitudes.} This is indeed a bit surprising,
since the Riemann identities typically mix {\it all} the internal and
space-time sectors (cf., \trialitymap).

However, our amplitudes do not (as usual) amount to fermion
number current insertions, but to more complicated fermionic
contractions, and the results of Appendix \appB\ show that, in the net
result, the elliptic genus of $K3$ happens to cancel out.
Heuristically one may say that this is because our correlators
probe only the $T^2$ sector of the theory because they involve only $T$
and $U$.

\chapter{Non-perturbative results for the N=4 heterotic string}

So far we have been dealing with perturbative quantities
in the type IIA string on $K3\times T^2$. The interesting issue
now is to map these via duality to the heterotic string on $T^6$,
by identifying \doubref\md\onequarterBPS:
$$
\eqalign{
T\ &=\ S_\het\ \equiv\ {\theta\over 2\pi}+{4\pi i\over
{g_\het}^2}\ \equiv\ a+i e^{-\Phi}
\cr
S\ &=\ T_\het\ \equiv\ B_{45}^\het+i \sqrt{|G^\het|}
\cr
 U\ &=\ U_\het\ \equiv (G_{45}^\het+i\sqrt{|G^\het|})/G_{44}^\het
\ ,}
\eqn\dualmap
$$
where $T_\het,U_\het$ correspond to the two-torus in $T^6=T^4\times
T^2$.

The perturbative $T$-dependence that we have been computing before will
thus give non-perturbative information about the heterotic string.
Remember that we have been suppressing non-perturbative corrections in
the type IIA string by going to weak coupling, by sending
$S=T_\het\to i\infty$. This corresponds to the decompactification limit
of the heterotic two-torus.

More specifically, while the Kaluza-Klein modes (labelled by $m_i$ in
\momenta) remain KK modes in the heterotic string,  the type IIA
windings around 1-cycles of $T^2$ (labelled by $n_i$)  turn into
magnetically charged  wrapping modes of the heterotic
fivebrane\foot{Since we are at a generic point in the Narain moduli
space, where there are no non-abelian gauge symmetries, this is the
neutral heterotic fivebrane \multref\soojong\MDXL\CHS\ with zero size,
or a ``small instanton'' \KthreeBPS.} around 5-cycles in $T^6$.
Alternatively, one may imagine wrapping the fivebrane first around the
sub-torus $T^4$, to yield a string in six dimensions that is dual to
the heterotic string \refs{\md{,} \SenII{,}\HS{,} \onequarterBPS{,}
\KthreeBPS}. The type IIA windings $n_i$ are then the same as the
windings of this dual string around 1-cycles of the remaining $T^2$ on
the heterotic side.  In total we thus have dyonic bound states of
wrapped fivebranes of charge $m_i$ with KK modes of momentum $n_j$,
which are 1/4 BPS if $m_1n_2-m_2n_1\not=0$ and 1/2 BPS if the DZW
product vanishes. The windings and momenta are exchanged by
$S$-duality, which is a non-perturbative symmetry from the heterotic
string point of view, but a perturbative one from either the type IIA
string or from the heterotic fivebrane point of view \doubref\SS\sen.

However, in analogy to the type IIA side, it is more natural to
interpret the prepotentials  in terms of instanton series. Quite
generally, world-sheet instantons are mapped under the duality to
space-time instantons, and indeed contributions of the form $e^{2 \pi
ik S_\het}$ correspond \harveymoore\ to gravitational fivebrane
instantons that arise from winding the heterotic fivebrane around the
whole of~$T^6$.

As far as the $U_\het$ dependence is concerned (which simply describes
KK excitations), it is actually more interesting to convert $U_\het\to
T_\het$, by making use of the $T_\het\--U_\het$ exchange symmetry of
the heterotic string. The purely $T_\het$ dependent terms then describe
heterotic world-sheet instanton contributions, and the mixed terms
in the prepotentials
$$
\eqalign{
\cG_1^\fk(S_\het,T_\het)
\attac{{harm.\atop piece}}\!\!\!\!\!\!\
&\sim
\Big({\del\over\del V}\Big)^{2K+2}
\widehat\cG_1(S_\het,T_\het,V)\attac{V\to0}
\cr
&\sim
\sum_{(k,l)>0}c^\fk(kl)\,
\Li_{1-2K}\lf(e^{2\pi ik S_\het} e^{2\pi il T_\het}\ri)
}\eqn\thefullthing
$$
must therefore be due to bound states or superpositions of fivebrane
instantons with world-sheet instantons.  In particular, the remarkable
factorization property \productform\ of the prepotential for the
six-point couplings,
$$
\eqalign{
\cG_1^{\scriptscriptstyle{[6]}}(S_\het,T_\het)
\attac{{holom.\atop piece}}\!\!\!\!\!\!\!\!\!\!
&= -{1\over24}\!\left(1\!-\!24\sum_{k>0} k {e^{2\pi i kS_\het}
\over 1\!-\!e^{2\pi i kS_\het}}\right)
\!\!
\left(1\!-\!24\sum_{l>0} l  {e^{2\pi i lT_\het}
\over 1\!-\!e^{2\pi i lT_\het}}\right)\!,\cr
}\eqn\STser
$$
tells us that the fivebrane and world-sheet instanton sectors that
contribute to these couplings must be essentially independent.

In fact, by investigating the
dependence on the coupling constants we find that the prepotentials
$\cG_1^\fk$ correspond to non-perturbative corrections to
$(2K\!+\!4)$-point amplitudes at $K$-loop order in the heterotic
string, so that the world-sheet instantons are of genus $g\leq
K$.\foot{The prepotentials $\cG_1^\fk$ are thus analogs of the
well-known prepotentials $F_g$  that arise (in $N=2$ supersymmetric
theories) at one-loop order in the heterotic string but at $g$ loops on
the type II side \refs{\bcov{,} \coniB{,}\AGNT{,}\AKEZ}.}

Note that world-sheet instantons on top of a fivebrane can also be
viewed as gauge instantons in the world-volume theory of the fivebranes
\MDBB. More specifically, it is known that a stack of $Q_5=k$ heterotic
fivebranes has $Sp(2k)$ gauge symmetry \doubref\KthreeBPS\MDGD.
Accordingly it has among other terms:
$$
\int d^6x\ \Big(\,B\wedge \mathop{\rm tr}_{Sp(2k)}F\wedge F
 +  {1\over {g_{5br}}^2}
\mathop{\rm tr}_{Sp(2k)}F^2\Big)
$$
on its world-volume, the first term being necessary for anomaly
cancellation \JM. It is known \doubref\md\onequarterBPS\  that the
``space-time'' coupling of the $T^4$-wrapped fivebrane (or dual string,
that is) is equal to the world-sheet coupling of the fundamental
string, which means: ${1\over {g_{5br}}^2}=\sqrt{|G^\het|}$. Comparing
to \dualmap, we thus see that a charge $Q_1=l$ instanton on top of a
charge $Q_5=k$ fivebrane will give an additional factor of $e^{2\pi i
lT_\het}$ besides $e^{2\pi i kS_\het}$, and this is what finally gives
a particularly interesting physical interpretation of the
$S_\het-T_\het$ mixing terms in \thefullthing. Moreover the instantons
must break one-half of the supersymmetries on the fivebrane, so that
the total configuration has only 1/4 unbroken supersymmetries.

Something non-trivial may then be learned for these gauge theories from
the numerical values of the coefficients of the mixing terms in
\thefullthing. In analogous situations such coefficients count either
isolated instantons, or Euler numbers of the moduli spaces if the
instantons are not isolated. Most likely the coefficients mean
something similar here too, and in particular $c^\fk(kl)$ should carry
information about the cohomology of moduli spaces of charge $l$, genus
$K$ instantons in the $Sp(2k)$ gauge theories on $T^6$ (the exchange
symmetry in $k$ and $l$  would relate this to charge $k$ instantons in
$Sp(2l)$ gauge theories). We hope to present a more complete discussion
elsewhere.

\ack

We would like to thank I.\ Antoniadis,  A.\ Brandhuber, B.\ de Wit, K.\
F\"orger, Y.~Oz, S.\ Rey,  A.\ Sen, H.\ Verlinde, N.P.\ Warner and in
particular J.\ Harvey, E.~Kiritsis  and P.\ Mayr for discussions. We
also would like to thank C.\ Vafa for valuable comments about the
manuscript. Moreover we thank the NATO Advanced Study Institute at
Carg\`ese for kind invitation and providing a pleasant atmosphere.

\appendix{\appA}{Generalized world--sheet torus integrals}

In this section we outline how to evaluate world--sheet torus integrals
of the following form:
$$
\eqalign{
&{\p^{q_1+q_2+q_3+q_4} \o \p\Lambda_1^{q_1} \p\Lambda_2^{q_2}
\p\Lambda_3^{q_3}
\p\Lambda_4^{q_4}}
\int {d^2\tau \o \tau_2}\ \tau_2^r  {1\o \tau_2^s}
\sum_{(p_L,p_R)}\ e^{\pi i\tau|p_L|^2} e^{-\pi i\ov\tau|p_R|^2}\cr
&\times \lf.
e^{\Lambda_1\ov p_R+ \Lambda_2p_R +\Lambda_3\ov p_L+\Lambda_4 p_L}\
e^{-{1\o
2\pi\tau_2}(\Lambda_1\Lambda_2+\Lambda_2\Lambda_3+\Lambda_3\Lambda_4+
\Lambda_4\Lambda_1)}\ f_{k}(\ov q)g_l(q)\ri|_{\Lambda_i=0}
\ ,\cr}
\eqn\todo
$$
where $f_k(\ov q)=\sum\limits_n c(n)\ov q^n$ and
$g_l(q)=\sum\limits_m d(m) q^m$ are modular functions of weights $k,l$,
respectively, and the integers obey $r,s\geq 0$,\  $q_1,q_2,q_3,q_4\geq
0$. Modular invariance of the integrand
requires:
$$
\eqalign{
q_1+q_2-r+s+k&=0\cr
q_3+q_4-r+s+l&=0\ .}
\eqn\conditions
$$
The  integral \todo\ can be performed with the
orbit decomposition method of \DKLII. In the following we will discuss
only the non--degenerate orbit $I_1$, as the degenerate orbit
has already been evaluated in ref.~\FS.

In the same reference, the non--degenerate orbit $I_1$ with $g_l(q)=1,
f_k(\ov q)\neq 1$ has been  worked out as well. A general feature of
$I_1$ is that the $T$ and $U$ moduli always appear in pairs in the
poly-logarithms that are either completely holomorphic
or anti-holomorphic, i.e., either $(T,U)$ and/or $(\ov T,\ov U)$
appear.

Before we go to the general case, let us discuss the example
$f_k(\ov q)=1, g_l(q)\neq 1$, which is in fact what we need in section
4:
$$
\eqalign{
&{\p^{q_1+q_2+q_3+q_4} \o \p\Lambda_1^{q_1}
\p\Lambda_2^{q_2} \p\Lambda_3^{q_3}
\p\Lambda_4^{q_4}}
\int {d^2\tau \o \tau_2}\ \tau_2^r  {1\o \tau_2^s}
\sum_{(p_L,p_R)}\ e^{\pi i\tau|p_L|^2} e^{-\pi i\ov\tau|p_R|^2}\cr
&\times \lf.
e^{\Lambda_1\ov p_R+ \Lambda_2p_R +\Lambda_3\ov p_L+\Lambda_4 p_L}\
e^{-{1\o 2\pi\tau_2}(\Lambda_1\Lambda_2+\Lambda_2
\Lambda_3+\Lambda_3\Lambda_4+
\Lambda_4\Lambda_1)}\ g_l(q)\ri|_{\Lambda_i=0}
\ .\cr}
\eqn\todoii
$$
The presence of a {\it holomorphic} function $g_l(q)$ (which is in
contrast to the usually considered situation) has as consequence that
now mixed holomorphic/anti-holomorphic pairs of  moduli appear in the
arguments of the poly-logarithms, i.e.,  $(T,\ov U)$ and/or $(\ov
T,U)$.
After introducing
$$\eqalign{
b&=p^2-
{i(\Lambda_1+\Lambda_2+\Lambda_3+\Lambda_4)p\o \pi \sqrt{2T_2U_2}}
-{1\o 8}{(\Lambda_1-\Lambda_2+\Lambda_3-\Lambda_4)^2\o \pi^2T_2U_2}\cr
\varphi&=p(kT_1+lU_1)\cr
&+{1\o 2\pi\sqrt{2T_2 U_2}}\lf[(kT_2-lU_2)\Lambda_1
+(-kT_2+lU_2)\Lambda_2+(-kT_2-lU_2)
\Lambda_3+(kT_2+lU_2)\Lambda_4\ri]\cr}
\eqn\short
$$
plus the function
$$
\tilde I_1(\alpha,\beta)=
{2 \o \sqrt{\beta b}}e^{-2\pi (kT_2-lU_2)\sqrt{\alpha\beta b}}
\ e^{-2\pi i\varphi}\ ,
\eqn\Ione
$$
we obtain the following closed expression for $I_1$ in the chamber
$T_2>U_2$:
$$\eqalign{
I_1&={\p^{q_1+q_2+q_3+q_4} \o \p\Lambda_1^{q_1} \p\Lambda_2^{q_2}
\p\Lambda_3^{q_3} \p\Lambda_4^{q_4}}\times\cr
&\times\sum_{{k>0 \atop l \in \IZ}}\sum_{p\neq 0}d(-kl)
\lf[{T_2U_2 \o \pi(kT_2-l U_2)^2}\ri]^r {1\o (\pi T_2 U_2 b)^s}
 {(-1)^{r+s}\p^{r+s}\o \p\alpha^r\p\beta^s}
\tilde I_1(\alpha,\beta)\lf.\ri
|_{{\alpha=1\atop \beta=1},\Lambda_i=0}\
.\cr}
\eqn\Itwo
$$
Finally, for the general case where $f_k,g_l\neq 1$, the
expressions \Ione\ and \Itwo\ are modified to
$$
I_1(\alpha,\beta)=
{2 \o \sqrt{\beta b}}e^{-2\pi \sqrt{(kT_2+lU_2)^2+4T_2U_2(n-kl)}
\sqrt{\alpha\beta b}}
\ e^{-2\pi i\varphi}
\eqn\Ionegeneral
$$
and
$$\eqalign{
I_1&={\p^{q_1+q_2+q_3+q_4} \o \p\Lambda_1^{q_1} \p\Lambda_2^{q_2}
\p\Lambda_3^{q_3} \p\Lambda_4^{q_4}}\sum_n\sum_{{k>0 \atop l \in \IZ}}
\sum_{p\neq 0}c(n)d(n-kl)\times\cr
&\times\lf[{T_2U_2 \o \pi[(kT_2+lU_2)^2+4T_2U_2(n-kl)]}\ri]^r
{1\o (\pi T_2 U_2 b)^s}
 {(-1)^{r+s}\p^{r+s}\o \p\alpha^r\p\beta^s}
I_1(\alpha,\beta)\lf.\ri|_{{\alpha=1\atop \beta=1},\Lambda_i=0}\ ,
\cr}\eqn\Itwo
$$
The last equation represents the result for the non--degenerate orbit
of \todo. We see that the amount of holomorphic/anti--holomorphic
mixing  is determined by $m=n-kl$ and is absent for $m=0$. This
reflects in our context that 1/4 BPS states can mix holomorphic and
anti-holomorphic sectors, in contrast to the 1/2 BPS states.

\appendix{\appB}{Fermionic contractions and bipartite Graphs}
\def\ss#1{{\scriptstyle{#1}}}

In this section we want to calculate the following correlator
of $2N$ real fermions (where $N\equiv(2K+2)$)
$$
\sum_{{{\rm spin}\atop {\rm structures}\  \a }}
\int dz_1\ldots\int dz_N\ \vev{\psi^{i_1}(z_1)\psi^{j_1}(z_1)\ldots
\psi^{i_N}(z_N)\psi^{j_N}(z_N)}_\al\ ,
\eqn\Dec
$$
which appears in the amplitudes
$\tilde{\cA}_{(\p_\nu T\p^\nu U)^{K+1} (\p_\rho \phi\p^\rho\ov
\phi)}^\fk$ in \higher.  For these amplitudes, where
the left--moving sector is saturated with four fermionic insertions,
the computations reduce to the right--moving part only; this is in line
with the considerations about elliptic genera of section 3, and of
course a reflection of the fact that only 1/4 BPS states contribute.

We therefore expect \Dec\ to be given by a ($2K+2$)--fold derivative of
an elliptic genus, and by considering modular invariance we see that
this genus should be a Jacobi form of weight $-2$. There are only two
natural candidates for it, namely either the total elliptic genus
$$
\eqalign{
\CHI(v_1,v_2,q)\ &=\
\CHI_{({\rm st}\times
T^2)}[\shalf(v_1+v_2),q]\ \CHI_{\kt}[\shalf(v_1-v_2),q]\cr
&=
\!{i\theta_1[\shalf(v_1+v_2),q]\over\eta^3(q)}
\!{i\theta_1[\shalf(v_1+v_2),q]\over\eta^3(q)}
\CHI_{(K3)}\lf[\shalf(v_1-v_2),q\ri]\ ,
}\eqn\totalgen
$$
or just only the weight $-2$ factor of it, which is 
$\CHI_{({\rm st}\times T^2)}$.

To fix this ambiguity, our strategy will be to do the fermion
contractions in \Dec\ for a few values of $K$, integrate these and sum
over the spin structures, and then compare the results with the two
candidate genera.

To deal with the fermionic contractions, we first decompose the
correlator \Dec\ into a product of  $N$ two--point functions:
$$
\vev{\psi^i(z_1)\psi^j(z_2)}_\alpha=
\delta^{ij}\ {\theta_\alpha(z_{12},\tau)\theta'_1(0,\tau)\o
\theta_\alpha(0,\tau)\theta_1(z_{12},\tau)}\ .
\eqn\corr
$$
There are in general many of such partitions, and denoting
contractions  $\vev{\psi^i(z_i)\psi^j(z_j)}$ by $(ij)$, the pattern
looks like:
$$\eqalign{
[2,2,2\dots,2,2] \ &\sim\ (1 2)^2(3 4)^2\ldots (N-1,{N})^2\ ,\cr
[2,2\dots,2,4] \ &\sim\ (12)(23) (3 4) (4 1)\times (5 6)^2\ldots
(N-1,N)^2\ , \cr
\vdots\ \ &\qquad \ \ \vdots\ \ \ \ \vdots\ \ \ \ \vdots\cr
[N]\ &\sim\ (12)(23)(34)\ldots (N1)\ .\cr}
\eqn\possibilities
$$
Note that these contractions form cycles, and the idea is to perform
the integrations and spin structure sums just for one (canonically
ordered) representative of each cycle class, $P$ (above, the cycle
class is indicated on the left). Indeed, as the ordering of the
positions $z_i$ influences the $z_i$--integrations in \Dec, each cycle
class in \possibilities\ will in general lead, after integration, to a
different modular function $g_\al^P(\bar q)$. Subsequently will then
need to multiply the result for each cycle class with the appropriate
combinatorial factor.

Moreover, we need to sum over the spin structures $\a$, which amounts
to folding the $g^P_\al(\bar q)$ into the right--moving part of the
partition function, which then yields new functions $G^P(\bar q)$. We
find it easiest to write this map in terms of a $\IZ_N$ orbifold limit
of the $K3$:
$$
G^{P}(\ov q)=\sum_{(h,g)\neq (0,0)}{\eta(\ov q)^{-6}\o
\Theta\lf[{1+h\atop 1+g}\ri]\Theta\lf[{1-h\atop 1-g}\ri]}
\sum_{(\alpha,\beta)}g^{P}_\al(\ov q)\
\Theta^2\lf[{\al\atop\beta}\ri](\ov q)
\Theta\lf[{\alpha-h\atop \beta-g}\ri](\ov q)
\Theta\lf[{\alpha+h\atop \beta+g}\ri](\ov q)\ ,
\eqn\gen
$$

Taking all this together, we see that \Dec\ may be written in the
following way:\foot{Correlators similar to \Dec\ appear in $N$ gauge
boson amplitudes with fermionized currents.  In that case the
combinations \possibilities\ correspond to the different invariants
$(\tr F^{2n})^m\ , 2nm=N$,  which arise in the decomposition of the $N$
gauge boson amplitude. For $N=4$ this has been studied before in the
literature \contr.}
$$\eqalign
{
\sum_\a\int dz_1\ldots\int dz_N &
\ \vev{\psi^{i_1}(z_1)\psi^{j_1}(z_1)\ldots
\psi^{i_N}(z_N)\psi^{j_N}(z_N)}_\al\cr
&=\ (2\tau_2)^{N-1}\pi^N\sum_{{{{\rm cycle}\atop{\rm classes\ }P}}} n_P
\ G^P(\bar q)\ ,\cr
}
\eqn\dec
$$
where $n_P$  denote the combinatorial factors
that count the number of permutations in \possibilities.

\noindent Let us now compute the (quasi)-modular forms
$G^{P}$ for some low values of $N\equiv 2K+2$:
\vskip .1cm\noindent{\it 4 fermion contractions ($N=2$):}
There is only the possibility $(12)^2$, which yields:
$$
g_\al^{[2]}(\tau)={1\o 3}(E_2+e_\al)\ ,
\eqn\Four
$$
where $e_\a(q)\ =-24q {d\o dq}\ln \theta_\al(q)-E_2(q)$. Inserted in
\gen, this gives\foot{We perform the calculation in an $\IZ_2$ orbifold
limit of $K3$. However the calculations can be easily generalized to
$\IZ_N$ orbifold limits along \stieberg.}:
$$
G^{[2]}=-12\ .
\eqn\Basic
$$
\vskip .1cm\noindent{\it 8 fermion contractions ($N=4$): }
There are two possibilities, namely, $(i)\  (12)^2(34)^2$ and $(ii)\
(12)(23)(34)(41)$ which integrate to
$$\eqalign{
g^{[2,2]}_\al(\tau)&={1\o 3^2}(E_2+e_\al)^2\ ,\cr
g^{[4]}_\al(\tau)&={1\o 9}(E_4-e_\al^2)\ ,\cr}
\eqn\Eight
$$
respectively.
Inserting into \gen\ leads to:\foot{We remark that while the last type
of
contractions leads to a vanishing result for $K3\times T^2$ vacua, it
gives a non-zero contribution in the corresponding computation in the
heterotic string  and so leads to additional kinematics.}
$$\eqalign{
G^{[2,2]}\ &= -8 E_2\ ,\cr
G^{[4]}\ &=\ 0\ .\cr}
\eqn\eight
$$
In the following we will explicitly display the spin-structure
dependent correlator only for the last combination in \possibilities,
namely $g^{[N]}_\a(q)$, as the others are combinations of correlators
with less fermions.
\vskip .1cm\noindent{\it 12 fermion contractions ($N=6$):}
The chain $(12)(23)(34)(45)(56)(61)$ gives after the $z_i$--integration
$$
g^{[6]}_\al(\tau)\ =\ {2\o 45}(E_6-E_4e_\al)\ .
\eqn\Twelve
$$
Altogether, after the orbifold sum we obtain:
$$
\eqalign{
G^{[2,2,2]}\ &=-4 (E_2^2+E_4)\ ,\cr
G^{[2,2,4]}\ &=-{8\o 3}E_4\ ,\cr
G^{[6]}\ &=-{8\o 5}E_4\ .\cr
}\eqn\twelve
$$
\vskip .1cm\noindent{\it 16 fermion contractions ($N=8$):}
Similarly as before:
$$
g^{[8]}_\al(\tau)\ =\
{2^8\o 5040}(-{1\o 16}E_4^2-{5\o 48} e_\al^4+
{1\o 6}E_4e_\al^2)\ ,
\eqn\Sixten
$$
and
$$
\eqalign{
\ss{(12)^2(34)^2(56)^2(78)^2}\ \ &\ \ \ss{G^{[2,2,2,2]}=-{8\o 9}
(2E_2^3+6E_2E_4+E_6)}\cr
\ss{(12)(23)(34)(41)\times (56)(67)(78)(85)}\ \ &\ \
\ss{G^{[4,4]}=-{8\o 9}E_6}\cr
\ss{(12)(23)(34)(45)(56)(61)\times (78)^2}\ \ &\ \
\ss{G^{[2,6]}=-{8\o 15}(E_6+E_2E_4)}\cr
\ss{(12)(23)(34)(41)\times  (56)^2(78)^2}\ \ &\ \
\ss{G^{[2,2,4]}=-{8\o 9}(E_6+2E_2E_4)}\cr
\ss{(12)(23)(34)(45)(56)(67)(78)(81)}\ \ &\ \
\ss{G^{[8]}=-{8\o 21}E_6}\ .
\cr}
\eqn\sixten
$$
\vskip .1cm\noindent{\it 20 fermion contractions ($N=10$):}
$$
g^{[10]}_\al(\tau)={-2^{10}\o 362880}({11\o 12}E_4E_6+
{5\o 6} E_6 e_\al^2+{7\o 4}E_4^2e_\al)\ .
\eqn\Twenty
$$
and
$$
\eqalign{
\ss{(12)^2(34)^2(56)^2(78)^2(9,10)^2}
\ \ &\ \ \ss{G^{[2,2,2,2,2]}=-{4\o 27}
(5E_2^4+30E_2^2E_4+10E_2E_6+9E_4^2)}\cr
\ss{(12)(23)(34)(41)\times (56)(67)(78)(89)(9,10)(10,5)}
\ \ &\ \ \ss{G^{[4,6]}=-{16\o 45}E_4^2}\cr
\ss{(12)(23)(34)(41)\times (56)(67)(78)(85)\times (9,10)^2}
\ \ &\ \ \ss{G^{[2,4,4]}=-{8\o 27}(2E_4^2+E_2E_6)}\cr
\ss{(12)(23)(34)(45)(56)(67)(78)(81)\times (9,10)^2}
\ \ &\ \  \ss{G^{[2,8]}=-{16\o 945}(18E_4^2+{15\o 2} E_2E_6)}\cr
\ss{(12)(23)(34)(45)(56)(61)\times (78)^2 (9,10)^2}
\ \ &\ \  \ss{G^{[2,2,6]}=-{8 \o 45}(3E_4^2+2E_2E_6+E_2^2E_4)}\cr
\ss{(12)(23)(34)(41)\times (56)^2(78)^2 (9,10)^2 }
\ \ &\ \  \ss{G^{[2,2,2,4]}=-{8\o 9}(E_4^2+E_2E_6+E_2^2E_4)}\cr
\ss{(12)(23)(34)(45)(56)(67)(78)(89)(9,10)(10,1)}
\ \ &\ \  \ss{G^{[2,2,2,2,2]}=-{8\o 45}E_4^2}\ .
\cr}
\eqn\twenty
$$
\vskip .3cm\noindent{\it 24 fermion contractions ($N=12$):}\br
$$
g^{[12]}_\al(\tau)={2\o 4455}(-{34\o 63} E_6^2-{1\o 5} E_4^3 +{92\o 63} E_6e_\al^3
+{11\o 5} E_4^2e_\al^2)\ ,
\eqn\Twentyfour
$$
and
$$\hskip-1cm\eqalign{
\ss{(12)^2(34)^2(56)^2(78)^2(9,10)^2(11,12)^2}
&\ss{\ \   G^{[2,2,2,2,2,2]}=-{1\o 27} (8E_2^5+80E_2^3E_4+40E_2^2E_6}
\cr &\qquad\qquad\qquad \ss{+72E_2E_4^2+16E_4E_6)}\cr
\ss{(12)(23)(34)(41)\times (56)(67)(78)(89)(9,10)(10,11)(11,12)(12,1)}
&\ss{\ \  G^{[4,8]}=-{136\o 945} E_4E_6} \cr
\ss{(12)(23)(34)(45)(56)(61)\times (78)(89)(9,10)(10,11)(11,12)(12,1)}
&\ss{\ \  G^{[6,6]}=-{32\o 225} E_4E_6} \cr
\ss{(12)(23)(34)(41)\times (56)(67)(78)(81)\times
(9,10)(10,11)(11,12)(12,1)}
&\ss{\ \  G^{[4,4,4]}=-{8\o 27} E_4E_6} \cr
\ss{(12)(23)(34)(41)\times (56)(67)(78)(85)\times (9,10)^2 (11,12)^2}
&\ss{\ \  G^{[2,2,4,4]}=-{8\o 81} (E_2^2E_6+4E_2E_4^2+4E_4E_6)}\cr
\ss{(12)(23)(34)(41)\times (56)(67)(78)(89)(9,10)(10,1)\times
(11,12)^2}
&\ss{\ \  G^{[2,4,6]}=-{16\o 135} (E_2E_4^2+2E_4E_6)}\cr
\ss{(12)(23)(34)(45)(56)(67)(78)(89)(9,10)(10,1)\times (11,12)^2}
&\ss{\ \  G^{[2,10]}=-{8\o 135} (E_2E_4^2+{41\o 21} E_4E_6)}\cr
\ss{(12)(23)(34)(45)(56)(67)(78)(81)\times (9,10)^2(11,12)^2}
&\ss{\ \  G^{[2,2,8]}=-{8\o 189} (E_2^2E_6+{24\o 5}E_2E_4^2
       +{22\o 5} E_4 E_6)}\cr
\ss{(12)(23)(34)(45)(56)(61)\times (78)^2(9,10)^2(11,12)^2}
&\ss{\ \  G^{[2,2,2,6]}=-{8\o 15}({1\o 9}E_2^3E_4+{1\o
 3}E_2^2E_6+E_2E_4^2+{5\o 9}E_4E_6)}\cr
\ss{(12)(23)(34)(41)\times (56)^2(78)^2(9,10)^2(11,12)^2}
&\ss{\ \  G^{[2,2,2,2,4]}=-{8\o 27}({4\o 3}E_2^3E_4+2E_2^2E_6+
     4E_2E_4^2+{5\o 3}E_4E_6)}\cr
\ss{(12)(23)(34)(45)(56)(67)(78)(89)(9,10)(10,11)(11,12)(12,1)}
&\ss{\ \  G^{[12]}=-{736\o 10395} E_4E_6}\ .
\cr}
\eqn\twentyfour
$$

In order to assemble these modular functions into \dec, we still need
to find the combinatorial factors $n_P$.  For this we employ a
graphical method, somewhat similar in spirit to that what was used in
the second reference of \contr. Indeed all the
contractions can be represented by graphs,  which can be labelled by
their cycle structure.

More precisely, we need to consider graphs with two kinds of vertices,
one kind referring to the moduli $T$ and the other to the moduli $U$.
That is, the first kind of vertices correspond to operators
$(k\cdot\psi)\Psi$, while the second kind corresponds to
$(k\cdot\psi)\bar\Psi$. Charge conservation for $\Psi,\bar\Psi$ and the
kinematical structure of the form $(\del_\mu T\del^\mu U)^{K+1}$ then
implies that only contractions between the two sets of vertices are
allowed; furthermore the contractions must form loops made from
alternating sequences of $\psi\bar\psi$ and $\Psi\bar\Psi$ propagators
-- see the figure.

\figinsert\graphs{
Bipartite graphs relevant for $K=2$. Each point on the left of a
diagram
corresponds to an operator $(k\cdot\psi)\Psi$, while on the right it
correspond to $(k\cdot\psi)\bar\Psi$. Each loop has to be
counted twice, reflecting the two ways to assign to it an alternating
sequence of $\psi\bar\psi$ and $\Psi\bar\Psi$ propagators. The cycle
structure also determines the signs.
 }{1.2in}{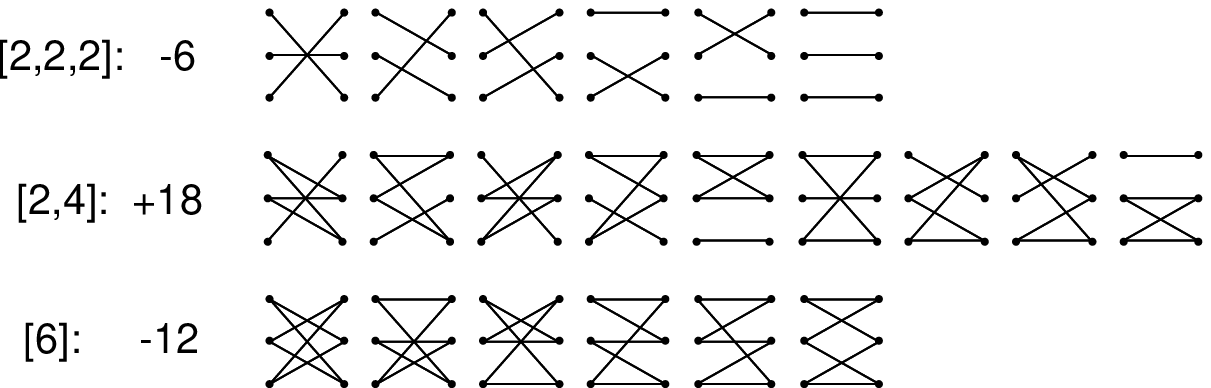}

Such ``bipartite'' graphs are characterized by incidence matrices of
the
block from
$$
I\ =\ \pmatrix{ 0 & R\cr R^t & 0\cr}
$$
where $R=\sum_{i,j}(r_i+r_j)$ and where $\{r_i\}$ are all the
permutations of the columns of the $N/2=K+1$ dimensional identity
matrix. Clearly there are in total $((K+1)!)^2$ such
graphs. Each of such graphs needs now to be classified with respect to
its cycle structure, which also determines its sign (given by the
signature of the permutation).

Being pragmatic, we generated these graphs up to $N=12$ with
Mathematica (which gives a formidable number), and decomposed them in
terms of their cycle structure.  In this way, we obtained the following
list of combinatorial coefficients:
$$
\eqalign{
N=4:\qquad
&{n_{[{2, 2}]}} = 2,{n_{[{4}]}} = -2
\cr
N=6:\qquad
&{{{n}}_{[{2, 2, 2}]}} = -6,\
{{{n}}_{[{2, 4}]}} = 18,\
{{{n}}_{[{6}]}} = -12
\cr
N=8:\qquad
&{n_{[{2, 2, 2, 2}]}} = 24,{n_{[{2, 2, 4}]}} = -144,\cr
&  {n_{[{4, 4}]}} = 72,{n_{[{2, 6}]}} = 192,{n_{[{8}]}} = -144
\cr
N=10:\qquad
&{n_{[{2, 2, 2, 2, 2}]}} = -120,{n_{[{2, 2, 2, 4}]}} = 1200,
  {n_{[{2, 4, 4}]}} = -1800,
\cr&{n_{[{2, 2, 6}]}} = -2400,
  {n_{[{4, 6}]}} = 2400,{n_{[{2, 8}]}} = 3600,
\cr&{n_{[{10}]}} = -2880
\cr
N=12:\qquad
&{n_{[{2, 2, 2, 2, 2, 2}]}} = 720,{n_{[{2, 2, 2, 2, 4}]}} = -10800,
  {n_{[{2, 2, 4, 4}]}} = 32400,
\cr&
{n_{[{2, 2, 2, 6}]}} = 28800,
  {n_{[{4, 4, 4}]}} = -10800,{n_{[{2, 4, 6}]}} = -86400,
\cr&
  {n_{[{2, 2, 8}]}} = -64800,{n_{[{6, 6}]}} = 28800,
  {n_{[{4, 8}]}} = 64800,
\cr&
{n_{[{2, 10}]}} = 103680,
  {n_{[{12}]}} = -86400.
}
$$
Inserting these together with our expressions for the $G^P(q)$ into
\dec, then produces combinations of Eisenstein series that {\it
exactly} match the following derivatives of $\CHI_{({\rm st}\times
T^2)}(v,q)$:
$$ \sum_{{{\rm cycle}\atop{\rm classes\ }P}}
n_{P(K)}G^P_{(K)}(\bar q)\ =\ \Big({\del\over\del v}\Big)^{2K+2}\!\!
\left(\!{i\theta_1(\shalf v,q)\over\eta^3(q)}\right)^2\attac{v=0}\ ,
$$
for $K=0,\dots,5$. By the uniqueness of the Jacobi forms this makes
clear what the relevant elliptic genus is, and in particular that the
elliptic genus of $K3$ cancels out.


\refout
\vfill
\eject
\end